\documentclass[twocolumn,english]{revtex4-2}
\usepackage{amsmath}
\usepackage{amssymb}
\usepackage[T1]{fontenc}   
\usepackage{mathptmx}      
\usepackage{graphicx}

\makeatletter

\providecommand{\tabularnewline}{\\}

\makeatother

\begin{document}
\title{Linear response of molecular polaritons}
\author{Joel Yuen-Zhou and Arghadip Koner}
\affiliation{Department of Chemistry and Biochemistry, University of California San Diego, La
Jolla, California 92093, USA}
\email{joelyuen@ucsd.edu, akoner@ucsd.edu} 
\begin{abstract}
In this article, we show that the collective light-matter strong coupling
regime, where $N$ molecular emitters couple to the photon mode of
an optical cavity, can be mapped to a quantum impurity model where
the photon is the impurity that is coupled to a bath of anharmonic
transitions. In the thermodynamic limit where $N\gg1$, we argue that
the bath can be replaced with an effective harmonic bath, leading
to a dramatic simplification of the problem into one of coupled harmonic
oscillators. We derive simple analytical expressions for linear optical
spectra (transmission, reflection, and absorption) where the only
molecular input required is the molecular linear susceptibility. This
formalism is applied to a series of illustrative examples showcasing
the role of temperature, disorder, vibronic coupling, and optical
saturation of the molecular ensemble, explaining that it is useful
even when describing an important class of nonlinear optical experiments.
For completeness, we provide a comprehensive Appendix that includes
a self-contained derivation of the relevant spectroscopic observables
for arbitrary anharmonic systems (for both large and small $N$) within
the rotating-wave approximation. While some of the presented results
herein have already been reported in the literature, we provide a
unified presentation of the results as well as new interpretations
that connect powerful concepts in open quantum systems and linear
response theory with molecular polaritonics.
\end{abstract}
\maketitle

\section{Introduction}

Polaritons are hybrid light-matter modes that emerge when the rate
of coherent energy exchange between material transitions and a confined
electromagnetic mode surpasses their respective dephasing rates. Hereafter,
we shall be concerned with molecular polaritons, where the material
transitions involve electronic and rovibrational states of molecules.
This light-matter strong coupling (SC) regime pertains to the interaction
of a large ensemble of $N\gg1$ molecules with a cavity mode \citep{raphaelrev2018}.
The latter requirement is a consequence of light-matter coupling per
molecule being negligible in standard Fabry-Perot microcavities, thus
necessitating a mesoscopic ensemble of transitions to collectively
give rise to SC. Such a restriction is lifted in special nanophotonic
environments, where SC with small $N$ has been achieved in various
platforms \citep{Chikkaraddy2016,Leng2018,Bitton2019,haugland2021intermolecular,koner2023path}.
Throughout this article we shall be concerned with the large $N$
case, unless otherwise stated.

Over the last decade, molecular microcavities have drawn attention
for their potential impact on altering rates and selectivity of chemical
reactions \citep{Hutchison2012,Thomas2019}, enhancing energy transfer
processes \citep{Coles2014,Zhong2017,delpo2021polariton}, and enabling
room-temperature polariton condensation \citep{kena2010room,plumhof2014room},
among other exciting phenomena. Linear spectroscopy as in the measurement
of linear transmission, absorption, and reflection, has been widely
used to characterize many of these experiments and to demonstrate
the onset of the SC regime \citep{Torma2015}. While a number of experimental
studies use classical optics (transfer matrix methods \citep{zhu1990vacuum,schubert1996polarization,yariv2007photonics})
to succesfully model their spectra \citep{xiang2021nonlinear,simpkins2023control,wright2023rovibrational,gunasekaran2023continuum},
a variety of quantum optical methods based on input-output theory
\citep{Gardiner1985,Ciuti2006,SteckBook,portolan2008nonequilibrium}
have also been used to simulate molecular polaritons featuring a simplified
energy level structure \citep{Li2018,Ribeiro2018vp,reitz2019langevin,kansanen2021polariton}.
The latter approach seems to be at odds with the standard philosophy
of computational molecular spectroscopy, where the linear spectrum
of a complex molecular system is obtained through the time-dependent
calculation of a dipole-dipole correlation function \citep{heller1981semiclassical,MukamelBook,TannorBook,heller2018semiclassical}.
In this article, we show that the calculation of a photon-photon correlation
function is enough to obtain all linear optical spectra of the molecular
microcavity. Moreover, we prove that in the $N\to\infty$ limit, these
spectra simplify dramatically and can be computed by direct input
of the molecular linear susceptibility (which can be obtained by Fourier
transformation of the dipole-dipole correlation function), as shown
in \citep{Cwik2016,lieberherr2023vibrational}, without running an
explicit calculation of $N$ molecules coupled to a photon mode. While
these expressions have already been featured in works by Keeling and
coworkers \citep{Cwik2016,zeb2018exact}, we clarify their assumptions
and applicability, and provide a new and enlightening interpretation
by regarding polaritons as quantum impurity models \citep{Makri1999}. 

This article is structured in the following way. Section \ref{sec:Molecular-polaritons-as}
provides the setup of the polariton Hamiltonian and shows how it can
be regarded as a quantum impurity model where the photon mode as the
impurity (the system) is coupled to a large bath of anharmonic molecular
transitions. In the limit of $N\to\infty$, the reduced dynamics of
the photon can be obtained by replacing this complex bath with a surrogate
one of harmonic modes, rendering the problem to one of coupled harmonic
oscillators. The spectral density of this surrogate bath turns out
to be proportional to the imaginary part of the linear susceptibility
of the bare molecular ensemble, $J^{\text{eff}}(\omega)\propto\Im\chi(\omega)$.
Importantly, this $\chi(\omega)$ does not need to obey Boltzmann
statistics for a given temperature, but can correspond to a nonequilibrium
stationary state, rendering our formalism applicable even to nonlinear
optical experiments where the latter are optically prepared. Section
\ref{sec:Linear-spectroscopy} outlines the expressions for the linear
spectra for arbitrary $N$ in terms of the photon Green's function
and, for $N\to\infty$ shows its explicit relation to $\chi(\omega)$.
Section \ref{sec:Examples} provides several illustrative examples
that are of interest to current polaritonics experiments, including
effects of optical saturation, disorder, vibronic coupling, and nonequilibrium
stationary states, which can all be treated with the same formalism.
Finally, the Appendix provides a self-contained derivation of (A)
input-output theory, (B) a derivation of spectroscopic observables
based on Kubo linear response and the expression of transmission in
the form of a “Landauer formula”, and (C) the simplification
of spectra in the harmonic limit ($N\to\infty$) using Heisenberg
equations of motion or, alternatively, using Kubo linear response. 

For a practical use of the results in the article, we recommend the
reader to directly refer to Table \ref{tab:formulas}, which shows
the direct relation between the polariton spectra and $\chi(\omega)$
when $N\to\infty$, and to work through some of the examples in Section
\ref{sec:Examples}. Readers who are interested in the theoretical
connections between quantum impurity models, open quantum systems,
and polaritons are welcome to read Section \ref{sec:Molecular-polaritons-as}.
The Appendix can be consulted to understand the derivations of the
spectroscopic formulas. 

\section{Molecular polaritons as quantum impurity problems\label{sec:Molecular-polaritons-as}}

In the collective SC regime, we are interested in a (harmonic) cavity
mode of frequency $\omega_{ph}$ coupled to $N$ non-interacting quantum
emitters. The Hamiltonian describing this setup (hereafter denoted
as the “molecular microcavity”) is 
\begin{equation}
H=H_{0}+V,\label{eq:H-1}
\end{equation}
where 
\begin{equation}
H_{0}=H_{ph}+H_{mol}\label{eq:H0-1}
\end{equation}
is the zeroth order contribution describing the photon and molecular
degrees of freedom,

\begin{align}
H_{ph} & =\ensuremath{\hbar\omega_{ph}}a^{\dagger}a,\label{eq:Hph}\\
H_{mol} & =\sum_{i}H_{i}(\boldsymbol{q}_{i},\boldsymbol{Q}_{i})\nonumber \\
 & =\sum_{y}\hbar\Omega_{y}|y\rangle\langle y|.\label{eq:Hmol}
\end{align}
Here $[a,a^{\dagger}]=1$ and the molecular term is a sum of contributions
of different molecules $i$, each of which depends on its respective
electronic and nuclear degrees of freedom $\boldsymbol{q}_{i}$, $\boldsymbol{Q}_{i}$.
We formally decompose $H_{mol}$ as a sum over many-body eigenstates
$|y\rangle$. Finally, for concreteness, the light-matter interaction
is taken to be dipolar,

\begin{equation}
V=-\hbar\lambda(a+a^{\dagger})\mu,\label{eq:V-1}
\end{equation}
where $\mu=\sum_{i}\mu_{i}(\boldsymbol{q}_{i},\boldsymbol{Q}_{i})$
is the dipole operator and $\hbar\lambda=\sqrt{\frac{\hbar\omega_{ph}}{2\epsilon_{0}\mathcal{V}}}$
is the vacuum electric field ($\epsilon_{0}$ is the permittivity
of vacuum and $\mathcal{V}$ is the cavity mode volume). Here, notice
that the “counterrotating” light-matter interaction terms are
present; the diamagnetic terms that are proportional to $(a+a^{\dagger})^{2}$
can always be removed via a Bogoliubov transformation \citep{Cwik2016}. 

At this point, it is conceptually convenient to invoke nomenclature
from the open quantum systems literature and regard the photon mode
as the \emph{system} and the molecular degrees of freedom as the \emph{bath}.
This scenario corresponds to the archetypal problem of an quantum
impurity coupled to a large enviroment (the standard collective strong-coupling
scenario consists of a large number of emitters $N\approx10^{3}-10^{10}$).
Makri \citep{Makri1999} has rigorously shown that if the environment
is constituted by an infinite number of independent degrees of freedom,
and if the system and the bath start in an uncorrelated state, 

\begin{align}
\rho(t_{\text{in}}) & =\rho_{ph}\otimes\rho_{mol}\nonumber \\
 & =\rho_{ph}\otimes\sum_{y}p_{y}|y\rangle\langle y|,\label{eq:product_state_section-1}
\end{align}
where we take $\rho_{mol}$ to be an arbitrary stationary density
matrix with respect to $H_{mol}$, with $p_{y}$ being the probability
of being in eigenstate $|y\rangle$ ( \citep{Makri1999} discusses
only the thermal case; the extension to arbitrary stationary states,
which is relevant in molecular polaritonics, is presented below).
The \emph{reduced system dynamics} can be obtained exactly using a
surrogate quantum impurity Hamiltonian $H^{\text{eff}}$,

\begin{equation}
H^{\text{eff}}=H_{0}^{\text{eff}}+V^{\text{eff}},\label{eq:H_eff}
\end{equation}
where we replace the original (in general, anharmonic) bath with a
surrogate harmonic bath,

\begin{align}
H_{mol}^{\text{eff}} & =\sum_{j}\hbar\omega_{j}b_{j}^{\dagger}b_{j},\label{eq:Heff_mol}
\end{align}
the system remains the same, $H_{ph}^{\text{eff}}=H_{ph}$, 

\begin{align}
H_{0}^{\text{eff}} & =H_{mol}^{\text{eff}}+H_{ph},\label{eq:H0_eff}
\end{align}
and the interaction between the system and the bath is,

\begin{align}
V^{\text{eff}} & =-(a+a^{\dagger})\sum_{j}c_{j}x_{j}\nonumber \\
 & =-(a+a^{\dagger})\sum_{j}\hbar\bar{c}_{j}(b_{j}+b_{j}^{\dagger}).\label{eq:Veff}
\end{align}
The couplings $\hbar\bar{c}_{j}=c_{j}\sqrt{\frac{\hbar}{2m_{j}\omega_{j}}}$
are characterized by an effective spectral density
\begin{align}
J^{\text{eff}}(\omega) & =\Theta(\omega)\pi\hbar\sum_{j}|\bar{c}_{j}|^{2}\delta(\omega-\omega_{j}).\label{eq:Jeff_definition}
\end{align}

\begin{figure*}
\begin{centering}
\includegraphics[scale=0.03]{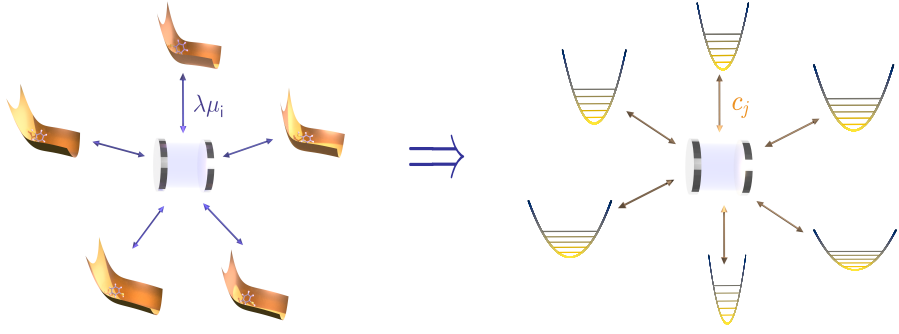}
\par\end{centering}
\caption{The molecular polariton problem, where the anharmonic degrees of freedom
of a large number $N$ of molecules are coupled to a single (harmonic)
photon mode can be regarded as a quantum impurity model. When $N\to\infty$,
the reduced dynamics of the photon can be computed exactly by replacing
the molecular degrees of freedom with a surrogate harmonic bath. \label{fig:Makri_mapping}}
\end{figure*}

Here, we adopt the convention that the spectral density is only finite
for positive frequencies {[}the step function $\Theta(\omega)=1$
for $\omega\geq0$ and $\Theta(\omega)=0$ otherwise{]}. This spectral
density is chosen so that the two-point correlation function of the
effective bath,
\begin{align}
C_{2}^{\text{eff}}(t) & =\sum_{j}|c_{j}|^{2}\langle e^{iH_{mol}^{\text{eff}}t/\hbar}x_{j}e^{-iH_{mol}^{\text{eff}}t/\hbar}x_{j}\rangle\label{eq:C_eff}
\end{align}
coincides with that of the original system, 
\begin{equation}
C_{2}^{\text{eff}}(t)=C_{2}(t)\label{eq:surrogate bath condition}
\end{equation}
for all $t\geq0$, where

\begin{align}
C_{2}(t) & =|\hbar\lambda|^{2}\langle e^{iH_{mol}t/\hbar}\mu e^{-iH_{mol}t/\hbar}\mu\rangle\nonumber \\
 & =|\hbar\lambda|^{2}\sum_{i}\langle e^{iH_{mol}t/\hbar}\mu_{i}(\boldsymbol{q}_{i},\boldsymbol{Q}_{i})e^{-iH_{mol}t/\hbar}\mu_{i}(\boldsymbol{q}_{i},\boldsymbol{Q}_{i})\rangle.\label{eq:C2}
\end{align}
The trace in Eq. \ref{eq:C2} is carried out with respect to $\rho_{mol}$. 

Let us now be more explicit about the evaluation of Eq. \ref{eq:C_eff}.
Its trace is carried out with respect to an initial stationary state
of harmonic oscillators with a \emph{frequency-dependent} inverse
temperature $\beta^{\text{eff}}(\omega)$. This complexity arises
from the arbitrariness of $\rho_{mol}$; in the special case when
$\rho_{mol}$ corresponds to a thermal state at inverse temperature
$\beta=(k_{B}\mathcal{T})^{-1}$, the surrogate setup is also thermal
at effective inverse temperature $\beta^{\text{eff}}(\omega)=\beta$\footnote{More precisely, $\rho_{mol}$ cannot correspond to a system with population
inversion, as a harmonic oscillator cannot reproduce $C_{2}(t)$ in
that case. Hence, this mapping only works if $\beta^{\text{eff}}(\omega)\geq0$.}. This initial state does not need to be explicitly specified, given
the well-known result connecting the two-point correlation function
and the spectral density for a bath of harmonic oscillators \citep{NitzanBook}, 

\begin{align}
C_{2}^{\text{eff}}(t) & =\frac{\hbar}{\pi}\int_{-\infty}^{\infty}d\omega'J^{\text{eff}}(\omega')\nonumber \\
 & \times\bigg[\coth\bigg(\frac{\beta^{\text{eff}}(\omega')\hbar\omega'}{2}\bigg)\cos(\omega't)-i\sin(\omega't)\bigg].\label{eq:C2_eff in terms of Jeff}
\end{align}
Computing its Fourier transforms (see convention in Eq. \ref{eq:FT})
at $\pm\omega$,

\begin{subequations}\label{eq:C2,+-}

\begin{align}
C_{2}^{\text{eff}}(\omega) & =-i\int_{-\infty}^{\infty}dte^{i\omega t}C_{2}^{\text{eff}}(t)\nonumber \\
 & =-i\hbar J^{\text{eff}}(\omega)\Bigg[\coth\bigg(\frac{\beta^{\text{eff}}(\omega)\hbar\omega}{2}\bigg)+1\Bigg],\label{eq:C_eff_omega}\\
C_{2}^{\text{eff}}(-\omega) & =-i\int_{-\infty}^{\infty}dte^{-i\omega t}C_{2}^{\text{eff}}(t)\nonumber \\
 & =-i\hbar J^{\text{eff}}(\omega)\Bigg[\coth\bigg(\frac{\beta^{\text{eff}}(\omega)\hbar\omega}{2}\bigg)-1\Bigg],\label{eq:C_eff_-omega}
\end{align}
\end{subequations}\noindent and using Eq. \ref{eq:surrogate bath condition},
we obtain 
\begin{equation}
\beta^{\text{eff}}(\omega)=\frac{1}{\hbar\omega}\text{ln}\frac{C_{2}(\omega)}{C_{2}(-\omega)}.\label{eq:effective_temperature}
\end{equation}
With $C_{2}^{\text{eff}}(t)$ and $\beta^{\text{eff}}(\omega)$ in
hand, Eq. \ref{eq:C2_eff in terms of Jeff} can be solved for $J^{\text{eff}}(\omega)$

\begin{align}
J^{\text{eff}}(\omega) & =\frac{2}{\hbar}\text{tanh}\frac{\hbar\omega\beta^{\text{eff}}(\omega)}{2}\int_{0}^{\infty}\Re C_{2}^{\text{eff}}(t)\text{cos}\omega tdt\nonumber \\
 & =\frac{i\Theta(\omega)}{\hbar}\int_{-\infty}^{\infty}C_{2}(t)\text{sin}\omega tdt.\label{eq:Jeff}
\end{align}
Thus, knowledge of $C_{2}(t)$ of the original bath (Eq. \ref{eq:C2})
together with the expression in Eq. \ref{eq:Jeff} give rise to $J^{\text{eff}}(\omega)$
which, we shall emphasize, varies as a function of $\rho_{mol}$.
It is well understood that knowledge of the spectral density alone
completely characterizes the reduced system dynamics if the bath is
composed of independent harmonic oscillators \citep{NitzanBook}.

The intuition behind this mapping is the following. In general, the
reduced system dynamics depends on $n$-point correlation functions
of the bath ($n\geq2$). However, Makri has shown that when the bath
is large enough ($N\gg1$), only the two-point correlation function
$C_{2}(t)$ becomes relevant {[}the $n>2$ such functions decay as
$O(N^{-1/2})$ if $c_{i}=O(N^{-1/2})$, which is what typically happens,
as we shall show by explicit examples in the next sections{]}. It
is also well known that a system coupled to a harmonic bath linearly
through each of its coordinates $x_{j}$ (see Eq. \ref{eq:Veff})
has vanishing such functions for $n>2$. This result is essentially
a consequence of the central limit theorem. Hence, we are entitled
to replace the original bath with a surrogate harmonic bath so long
as the two-point correlation functions coincide. This mapping is useful
given that a plethora of tools to solve for the reduced dynamics of
systems coupled to harmonic baths have been developed over the last
decades.

Note that Eq. \ref{eq:C2} is just the usual dipole-dipole correlation
function, which can be re-expressed in terms of eigenstates of $H_{mol},$

\begin{align}
C_{2}(t)= & \text{lim}_{\gamma\to0^{+}}\sum_{y,z}\big\{ p_{y}\nonumber \\
 & \times|\hbar\lambda\langle z|\mu|y\rangle|^{2}e^{-i(\omega_{zy}-i\frac{\gamma}{2})t}\big\},\label{eq:C2_spectral_decomposition-1}
\end{align}
where $\omega_{zy}=\frac{E_{z}-E_{y}}{\hbar}$. Inserting Eq. \ref{eq:C2_spectral_decomposition-1}
into Eq. \ref{eq:Jeff} gives rise to

\begin{align}
J^{\text{eff}}(\omega)= & \Theta(\omega)\pi\hbar\sum_{y,z}(p_{y}-p_{z})|\lambda\langle z|\mu|y\rangle|^{2}\delta(\omega_{zy}-\omega)\nonumber \\
= & \Theta(\omega)\hbar\Im\chi(\omega),\label{eq:Jeff in terms of chi}
\end{align}
where \begin{subequations}\label{eq:chi}

\begin{align}
\chi(\omega)= & -\text{lim}_{\gamma\to0^{+}}\sum_{j}\frac{|\bar{c}_{j}|^{2}}{\omega-\omega_{j}+i\frac{\gamma}{2}}\label{eq:chi_1}\\
= & -\text{lim}_{\gamma\to0^{+}}\sum_{y,z}[p_{y}-p_{z}]\frac{|\lambda\langle z|\mu|y\rangle|^{2}}{\omega-\omega_{zy}+i\frac{\gamma}{2}}\label{eq:chi_2}
\end{align}
\end{subequations}\noindent Here, $\chi(\omega)$ is the molecular
linear susceptibility \citep{MukamelBook} generalized to arbitrary
initial stationary states, where we emphasize that $|y\rangle$ and
$|z\rangle$ are eigenstates of the entire molecular ensemble. Thus,
we have reached the physically appealing conclusion that the effective
spectral density of the bath that couples to the photon mode is simply
the absorption spectrum {[}$\Im\chi(\omega)${]} of the original molecular
ensemble\footnote{In the special case that all the $N$ molecules are identical, 
\[
\chi(\omega)=-\lim_{\gamma\to 0^{+}}\frac{1}{\hbar}\Sigma_{a,b}(p_{a}-p_{b})\frac{N\frac{\omega_{ph}}{2\epsilon_{0}\mathcal{V}}|\langle a|\mu|b\rangle|^{2}}{\omega-\omega_{ba}+i\frac{\gamma}{2}},
\]
where $a,b$ label single-molecule eigenstates, and we used $|\lambda|^{2}=\frac{\omega_{ph}}{2\hbar\epsilon_{0}\mathcal{V}}$,
where $\mathcal{V}$ is the cavity mode volume. Thus $\chi(\omega)$
is proportional to molecular concentration $\rho_{0}=\frac{N}{\mathcal{V}}$.
This expression coincides with the textbook expression in \citep{MukamelBook}
(see Eqs. 6.5a and 6.8b) except for a convenient factor of $\frac{\omega_{ph}}{2\epsilon_{0}}$
that simplifies the notation for our calculations for molecular microcavities.}. As we shall see below, for our purposes of modelling linear spectra
of polaritons, it is more convenient to work with $\chi(\omega)$
directly than with $J^{\text{eff}}(\beta,\omega)$. However, the former
can be readily obtained from the latter; comparing Eqs. \ref{eq:Jeff_definition}
and \ref{eq:chi}, we have that for $\omega\geq0$,

\begin{align}
\chi(\omega) & =-\text{lim}_{\gamma\to0^{+}}\frac{1}{\pi}\int_{-\infty}^{\infty}d\omega'\frac{J^{\text{eff}}(\omega')}{\omega-\omega'+i\frac{\gamma}{2}},\label{eq:chi_in_terms_of_J-1}
\end{align}
while for $\omega<0$ we can use $\chi(-\omega)=\Re\chi(\omega)-i\Im\chi(\omega)$.
Alternatively, $\chi(\omega)$ can be obtained from a Fourier transform
of the dipole correlation function in Eq. \ref{eq:C2} as \citep{MukamelBook},

\[
\chi(\omega)=-\frac{1}{\hbar^{2}}\text{lim}_{\gamma\to0^{+}}[C(\omega)+C^{*}(-\omega)]
\]
where we use the following Fourier transform convention (the $-i$
is added to be consistent with a standard convention of Green's function
theory, which will be invoked later),

\begin{subequations}\label{eq:FT}

\begin{align}
f(\omega) & =-i\int_{-\infty}^{\infty}dte^{i\omega t}f(t),\label{eq:x(w)}\\
f(t) & =\frac{i}{2\pi}\int_{-\infty}^{\infty}d\omega e^{-i\omega t}f(\omega).\label{eq:x(t)}
\end{align}
\end{subequations}

\section{Linear spectroscopy\label{sec:Linear-spectroscopy}}

\subsection{Formulas for arbitrary $N$\label{subsec:Formulas-for-arbitrary}}

The spectroscopic signals of cavity polaritons can be rigorously obtained
via input-output theory \citep{Gardiner1985,Ciuti2006,SteckBook,portolan2008nonequilibrium}.
For simplicity, we provide expressions under the rotating-wave approximation
(RWA) for light-matter interaction (see Eq. \ref{eq:V-1}), thus postponing
discussion of ultrastrong coupling systems for another study (at the
level of the RWA, the physics is already rich enough),

\begin{align}
V & =-\lambda a\mu^{>}-\lambda a^{\dagger}\mu^{<},\label{eq:RWA}
\end{align}
where $\mu^{>}=\sum_{\omega_{zy}>0}\langle z|\mu|y\rangle|z\rangle\langle y|$
is the dipole operator projected onto uphill transitions, and $\mu^{<}=\sum_{\omega_{zy}<0}\langle z|\mu|y\rangle|z\rangle\langle y|$
is the corresponding projector onto downhill transitions\footnote{Under moderate collective light-matter interaction couplings, if hereafter
Eq. \ref{eq:Veff} were to be used instead of Eq. \ref{eq:Veff-RWA},
the errors would be negligible given how off-resonant the non-RWA
terms are.}. 

The central quantity to compute is the photon retarded Green function,
\begin{align}
D^{R}(\omega) & =-i\int_{-\infty}^{\infty}dte^{i\omega t}\Theta(t)\langle e^{iHt/\hbar}ae^{-iH't/\hbar}a^{\dagger}\rangle,\label{eq:D_R(omega)_simple-main}
\end{align}
where 

\begin{equation}
H'=H-i\frac{\hbar\kappa}{2}a^{\dagger}a\label{eq:non-Hermitian-1}
\end{equation}
is the effective non-Hermitian Hamiltonian which accounts for coupling
of the cavity photon with the left and right photon continua, with
$\kappa=\kappa_{L}+\kappa_{R}$, and $\kappa_{L}$ and $\kappa_{R}$
labeling the corresponding rates of photon escape (hereafter, we use
the notation $X'=X-i\frac{\hbar\kappa}{2}a^{\dagger}a$ for any operator
$X$). Note that the first “$H$” in Eq. \ref{eq:D_R(omega)_simple-main}
is not primed while the second one is (see explanation around Eq.
\ref{eq:D_R(omega)_simple}). Eq. \ref{eq:D_R(omega)_simple-main}
yields frequency resolved transmission, reflection, and absorption
spectra, \begin{subequations}\label{eq:final_formulas-main}

\begin{align}
T(\omega) & =\kappa_{L}\kappa_{R}|D^{R}(\omega)|^{2},\label{eq:T_w-1}\\
R(\omega) & =1+2\kappa_{L}\Im D^{R}(\omega)+\kappa_{L}^{2}|D^{R}(\omega)|^{2},\label{eq:R_w-1}\\
A(\omega) & =-\kappa_{L}[\kappa|D^{R}(\omega)|^{2}+2\Im D^{R}(\omega)].\label{eq:A_w-1}
\end{align}

\end{subequations}\noindent 

\begin{figure}
\begin{centering}
\includegraphics[scale=0.06]{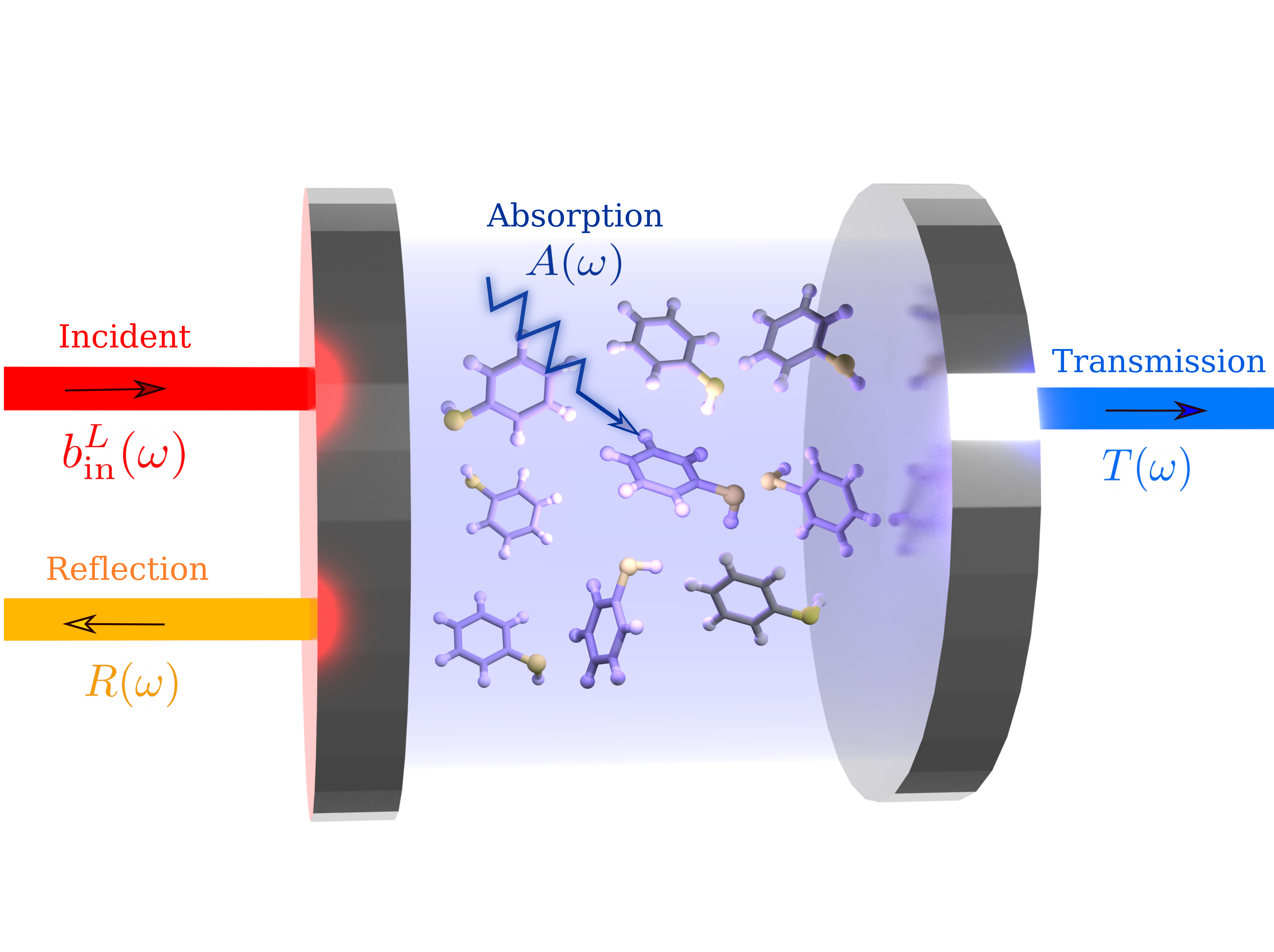}
\par\end{centering}
\caption{Linear spectroscopy of molecular polaritons as transmission, absorption,
and reflection (see Eq. \ref{eq:final_formulas-main}). \label{fig:spectroscopy}}
\end{figure}
Unlike most quantum optics treatments in the literature which deal
with $T=0$ and harmonic degrees of freedom alone, Eqs. \ref{eq:D_R(omega)_simple-main}
and \ref{eq:final_formulas-main} are valid for molecular microcavity
Hamiltonians $H$ (Eq. \ref{eq:H-1}) with arbitrary anharmonic molecular
degrees of freedom $H_{mol}$, arbitrary number of molecules $N$,
and the trace in Eq. \ref{eq:D_R(omega)_simple-main} is carried out
with respect to the initial density matrix $\rho_{0}$ of the molecular
microcavity, $\langle\cdot\rangle=\text{Tr}(\rho_{0}\cdot)$ which
is assumed to stationary under evolution with $H'$.

Eqs. \ref{eq:D_R(omega)_simple-main} and \ref{eq:final_formulas-main}
have been reported in the literature (see \citep{zeb2018exact}, Eqs.
15 and 16), although as far as we are aware, their explicit derivation
has not been presented anywhere (see also Eq. A4 in \citep{Cwik2016}
although it has a typo of $\frac{1}{2}$). For completeness, Appendix
\ref{sec:Formulas-for-linear} provides a self-contained derivation
of the latter, listing the assumptions involved.

\subsection{Formulas for $N\to\infty$}

Since the retarded photon Green function (Eq. \ref{eq:D_R(omega)_simple-main})
depends only on the photon \emph{system} (and not on the molecular
\emph{bath} degrees of freedom), we can capitalize on the statements
of Section \ref{sec:Molecular-polaritons-as}. Replacing $H\to H^{\text{eff}}$,
and working within the RWA, Eq. \ref{eq:Veff} becomes

\begin{align}
V^{\text{eff}} & =-a\sum_{j}c_{j}\sqrt{\frac{\hbar}{2m_{j}\omega_{j}}}b_{j}^{\dagger}+\text{h.c.}\label{eq:Veff-RWA}
\end{align}
An analytical expression for $D^{R}(\omega)$ can be obtained for
this harmonic surrogate setup; subsection \ref{subsec:effective harmonic}
(Appendix) presents explicit derivations. However, to provide additional
intuition, here we provide an alternative, possibly simpler derivation.
We exploit the profound fact that the response of a harmonic system
is \emph{independent} of initial condition of the oscillators (see
discussion in \ref{subsub EoM}, Appendix). This means that we might
as well take the initial state to be the vaccum $|0\rangle$ of all
the effective oscillators and the photon (even if $\rho_{mol}$ does
not correspond to $T=0$!). Then $H^{\text{eff}}|0\rangle=0$, so
Eq. \ref{eq:D_R(omega)_simple-main} becomes

\begin{align}
D^{R}(\omega) & =-i\int_{-\infty}^{\infty}dte^{i\omega t}\Theta(t)\langle0|ae^{-iH^{\text{eff}\prime}t/\hbar}a^{\dagger}|0\rangle\nonumber \\
 & =\langle1_{ph}|G^{\text{eff}}(\omega)|1_{ph}\rangle,\label{eq:DR_omega_as_simple_G}
\end{align}
where $|1_{ph}\rangle=a^{\dagger}|0\rangle$ is the one-photon state,
and

\begin{align}
G^{\text{eff}}(\omega) & =-i\int_{-\infty}^{\infty}dte^{i\omega t}\Theta(t)e^{-iH^{\text{eff}\prime}t/\hbar}\nonumber \\
 & =\frac{1}{\omega-H^{\text{eff}\prime}/\hbar}\label{eq:simple_G}
\end{align}
is the retarded Green function for $H^{\text{eff}\prime}$. Notice
that according to Eq. \ref{eq:Veff-RWA}, the light-matter coupling
$V^{\text{eff}}$ only mixes $|1_{ph}\rangle$ with states in the
first-excitation manifold of the effective harmonic bath $\{|1_{b_{j}}\rangle=b_{j}^{\dagger}|0\rangle\}$.
Thus, within the first excitation manifold, in the basis $\{|1_{ph}\rangle,\{|1_{b_{j}}\rangle\}\}$,
$H^{\text{eff}\prime}$ is an arrowhead matrix (that is, $|1_{ph}\rangle$
couples to the first-excitation states $|1_{b_{j}}\rangle$ but there
are no couplings among the latter or between the latter and other
states. Defining the zeroth-order retarded Green function $G_{0}^{\text{eff}}=[\omega-H_{0}^{\text{eff}\prime}/\hbar]^{-1}$,
we can deploy standard Green function identities,
\begin{align}
G_{0}^{-1} & =G^{-1}+\frac{V}{\hbar}\nonumber \\
\implies G_{0}G_{0}^{-1}G & =G_{0}G^{-1}G+G_{0}\frac{V}{\hbar}G\nonumber \\
\implies G & =G_{0}+G_{0}\frac{V}{\hbar}G\nonumber \\
 & =G_{0}+G_{0}\frac{V}{\hbar}G_{0}+G_{0}\frac{V}{\hbar}G_{0}\frac{V}{\hbar}G,\label{eq:G=00003DG0+...}
\end{align}

to obtain the desired expression \footnote{See for example, Eqs. 2.91 and 2.92 in \citep{MukamelBook} or Eq.
9.16 in \citep{NitzanBook}},

\begin{align}
D^{(R)}(\omega)= & \langle1_{ph}|G^{\text{eff}}(\omega)|1_{ph}\rangle\nonumber \\
= & \frac{1}{(\omega-\omega_{ph}+i\frac{\kappa}{2})-\Sigma_{M}},\label{eq:D(R)(E)}
\end{align}
where $\Sigma_{M}$ is the molecular self-energy, which using Eqs.
\ref{eq:Veff-RWA} and \ref{eq:Jeff in terms of chi} is found to
be minus the molecular susceptibility! 

\begin{align}
\Sigma_{M} & =\text{lim}_{\gamma\to0}\sum_{j}\frac{|\frac{1}{\hbar}V_{1_{ph},1_{j}}^{\text{eff}}|^{2}}{\omega-\omega_{j}+i\frac{\gamma}{2}}\nonumber \\
 & =\text{lim}_{\gamma\to0^{+}}\sum_{j}\frac{|\bar{c}_{j}|^{2}}{\omega-\omega_{j}+i\frac{\gamma}{2}}\nonumber \\
 & =-\chi(\omega).\label{eq:self_energy_as_minus_susceptibility_main}
\end{align}
Eqs. \ref{eq:D(R)(E)} and \ref{eq:self_energy_as_minus_susceptibility_main}
can be plugged into Eq. \ref{eq:final_formulas-main} to give the
compact formulas, \begin{subequations}\label{eq:spectroscopic_observables_harmonic-main}
\begin{align}
T(\omega)= & \frac{\kappa_{L}\kappa_{R}}{|\omega-\omega_{c}+i\frac{\kappa}{2}+\chi(\omega)|^{2}},\label{eq:T(w) harmonic}\\
A(\omega)= & \frac{2\kappa_{L}\Im\chi(\omega)}{|\omega-\omega_{c}+i\frac{\kappa}{2}+\chi(\omega)|^{2}},\label{eq:A(w) harmonic}
\end{align}
\end{subequations}\noindent and reflection can be easiest obtained
by subtraction, $R(\omega)=1-T(\omega)-A(\omega)$. Eq. \ref{eq:spectroscopic_observables_harmonic-main}
is the main result of the article. Incidentally, note that we may
write the transmission spectrum as a Landauer formula (see for example,
Appendix 9B in \citep{NitzanBook}, 

\begin{equation}
T(\omega)=\text{Tr}[\Gamma^{(L)}G^{\text{eff}\dagger}(\omega)\Gamma^{(R)}G^{\text{eff}}(\omega)],\label{eq:Landauer}
\end{equation}
where $\Gamma^{(i)}=\kappa_{i}|1_{ph}\rangle\langle1_{ph}|$ ($i=L,R$).
These results state that in the thermodynamic limit ($N\gg1$), information
about the molecular linear susceptibility alone is enough to predict
linear response properties of molecular microcavities, thus bypassing
simulations where the cavity is explicitly included. Importantly,
they have been derived assuming an initially uncorrelated state between
light and matter \ref{eq:product_state_section-1}, so they might
only be useful when SC happens for high-frequency molecular vibrations
\citep{Long2015,Shalabney2015} and electronic excitations \citep{Agranovich2003},
but shall be used with caution for collective low-frequency modes,
as in Terahertz polaritonics \citep{stoyanov2002terahertz,damari2019strong},
since finite temperature might populate polariton states and create
correlations between light and matter. As far as we are aware, these
statements were first made in \citep{Cwik2016,zeb2018exact}, and
are consistent with similar claims in \citep{perez2023simulating,lieberherr2023vibrational}.
Table \ref{tab:formulas} summarizes the main results of this article.

\begin{widetext}

\begin{table}
\begin{centering}
\begin{tabular}{|c|c|c|}
\hline 
\multicolumn{3}{|c|}{TABLE 1: Linear optics of molecular polaritons}\tabularnewline
\hline 
 & arbitrary $N$ & $N\to\infty$\tabularnewline
\hline 
\hline 
transmission & $T(\omega)=\kappa_{L}\kappa_{R}|D^{R}(\omega)|^{2}$ & $T(\omega)=\frac{\kappa_{L}\kappa_{R}}{|\omega-\omega_{c}+i\frac{\kappa}{2}+\chi(\omega)|^{2}}$\tabularnewline
\hline 
reflection & $A(\omega)=-\kappa_{L}[\kappa|D^{R}(\omega)|^{2}+2\Im D^{R}(\omega)]$ & $A(\omega)=\frac{2\kappa_{L}\Im\chi(\omega)}{|\omega-\omega_{c}+i\frac{\kappa}{2}+\chi(\omega)|^{2}}$\tabularnewline
\hline 
absorption & $R(\omega)=1+2\kappa_{L}\Im D^{R}(\omega)+\kappa_{L}^{2}|D^{R}(\omega)|^{2}$ & $R(\omega)=\frac{|\omega-\omega_{c}+i\frac{\kappa}{2}+\chi(\omega)|^{2}-\kappa_{L}\kappa_{R}-2\kappa_{L}\Im\chi(\omega)}{|\omega-\omega_{c}+i\frac{\kappa}{2}+\chi(\omega)|^{2}}$\tabularnewline
\hline 
\end{tabular}
\par\end{centering}
\caption{\label{tab:formulas}}
\end{table}

\end{widetext}

\section{Examples\label{sec:Examples}}

In the next subsections we will illustrate the formalism above, and
in particular, the use of the formulas in the rightmost column of
Table \ref{tab:formulas} in concrete examples highlighting effects
of temperature, disorder, and optical saturation of the molecular
ensemble. Even though the examples are simplified, they should serve
as pedagogical tools, highlighting the essence of these effects in
the spectra of polaritons. These examples can be readily generalized
to account for more realistic molecular details. 

\subsection{$N\gg1$ Two-level systems\label{subsec:-Two-level-systems}}

We start with the simple example of an ensemble of $N$ two-level
atoms coupled to the photon mode via their optical transition (Tavis-Cummings
model \citep{tavis1968exact}), \begin{subequations}\label{eq:Tavis-Cummings}

\begin{align}
H_{mol} & =\sum_{i=1}^{N}\hbar\omega_{exc,i}\sigma_{i}^{\dagger}\sigma_{i},\label{eq:Hmol_TC}\\
V & =-\hbar\lambda a\sum_{i=1}^{N}\mu_{i}\sigma_{i}^{\dagger}+\text{h.c.},\label{eq:V_TC}
\end{align}
\end{subequations}\noindent where $\sigma^{\dagger}=|e\rangle\langle g|$
and $\mu_{i}$ is the amplitude of the optical transition. The molecular
susceptibility according to Eq. \ref{eq:chi} is,

\begin{align}
\chi(\omega) & =-\sum_{i}\text{tanh}\frac{\beta\hbar\omega_{exc,i}}{2}\Bigg(\frac{|\lambda\mu_{i}|^{2}}{\omega-\omega_{exc,i}+i\frac{\gamma}{2}}\Bigg).\label{eq:chi_TC_no_disorder}
\end{align}
Let us first assume that all atoms are identical, $\omega_{exc,i}=\omega_{exc}$
and $\mu_{i}=\mu$. Denoting $g^{2}=|\lambda\mu|^{2}\text{tanh}\frac{\beta\hbar\omega_{exc}}{2}$,
it follows that $\chi(\omega)=-\frac{N|g|^{2}}{\omega-\omega_{exc,i}+i\frac{\gamma}{2}}$,
which can be inserted into Eq. \ref{eq:spectroscopic_observables_harmonic-main}
to give (see Fig. \ref{fig:Transmission,-reflection,-and} a), \begin{subequations}\label{eq:T,A}

\begin{align}
T(\omega) & =\frac{\kappa_{L}\kappa_{R}[(\omega_{0}-\omega_{exc})^{2}+(\frac{\gamma}{2})^{2}]}{|(\omega-\omega_{ph}+i\frac{\kappa}{2})(\omega-\omega_{exc}+i\frac{\gamma}{2})-Ng^{2}|^{2}},\label{eq:T(E)_TLS}\\
A(E) & =\frac{\kappa_{L}\gamma Ng^{2}}{|(\omega_{0}-\omega_{ph}+i\frac{\kappa}{2})(\omega_{0}-\omega_{exc}+i\frac{\gamma}{2})-Ng^{2}|^{2}}.\label{eq:A(E)_TLS}
\end{align}
\end{subequations}\noindent Eq. \ref{eq:T,A} are the standard results
for the simplest polariton system with collective light-matter coupling
is equal to $\sqrt{N}g$ (see for instance, the Supplementary Information
of \citep{Ribeiro2018vp}).

The renormalization of light-matter coupling $g$ with $\mathcal{T}$
has been previously obtained in \citep{Cwik2016}. Essentially, owing
to optical saturation, Rabi splitting contraction (also known as “phase
space filling” in the solid state literature \citep{deng2002condensation})
is expected for high temperatures as compared to $\mathcal{T}\to0$,
(see Fig. \ref{fig:Transmission,-reflection,-and}b). As $\mathcal{T}\to\infty$,
the light-matter coupling $g$ vanishes because the photon mode coherently
adds up the absorption and emission events; neither outside nor inside
the cavity do we expect any net absorption. In other words, this transparent
molecular medium inside a cavity makes the latter behave as an empty
cavity. These effects can be obtained in transient absorption experiments,
when optical pumping and dephasing ensues, creating steady-state populations
in excited molecular states \citep{Dunkelberger2016,xiang2019manipulating};
for instance, the $\mathcal{T}\to\infty$ was achieved in \citep{dunkelberger2019saturable}.
Thus, despite the nonlinear optical nature of transient absorption
experiments, we see that the (linear) transmission of the probe upon
dephasing and relaxation to the “dark states” obeys the very
simple physics outlined in this article, in agreement with the conclusions
in \citep{Ribeiro2018vp}. 

It is also instructive to rederive these results in a more pedestrian
way. Solving for the parameters of the surrogate setup,

\begin{subequations}\label{eq:surrogate_TLS} 

\begin{align}
J^{\text{eff}}(\beta,\omega) & =\hbar N|g|^{2}\delta(\omega_{exc}-\omega),\label{eq:Jeff_TLS}\\
H_{mol}^{\text{eff}} & =\hbar\omega_{exc}b^{\dagger}b,\label{eq:H_mol_eff_TLS}\\
V^{\text{eff}} & =-\sqrt{N}|g|(ab^{\dagger}+\text{h.c.}),\label{eq:Veff_TLS}
\end{align}

\end{subequations}\noindent In the \{$|1_{ph}\rangle,|1_{b}\rangle$\}
photon-exciton basis, we can write the effective Green function,

\begin{align}
G^{\text{eff}}(\omega) & =\Bigg[\begin{array}{cc}
\omega-\omega_{ph}+i\frac{\kappa}{2} & \sqrt{N}g\\
\sqrt{N}g & \omega-\omega_{exc}+i\frac{\gamma}{2}
\end{array}\Bigg]^{-1}\nonumber \\
 & =\frac{\Bigg[\begin{array}{cc}
\omega-\omega_{exc}+i\frac{\gamma}{2} & -\sqrt{N}g\\
-\sqrt{N}g & \omega-\omega_{ph}+i\frac{\kappa}{2}
\end{array}\Bigg]}{(\omega-\omega_{ph}+i\frac{\kappa}{2})(\omega-\omega_{exc}+i\frac{\gamma}{2})-\sqrt{N}g^{2}},\label{eq:G(E)_TLS}
\end{align}
as well as the matrices denoting coupling to photonic continua, $\Gamma^{(L)}=\left[\begin{array}{cc}
\kappa_{L} & 0\\
0 & 0
\end{array}\right]$, $\Gamma^{(R)}=\left[\begin{array}{cc}
\kappa_{R} & 0\\
0 & 0
\end{array}\right]$. Applying the Landauer-type formula of Eq. \ref{eq:Landauer}, we
readily recover Eq. \ref{eq:T(E)_TLS}.

\begin{figure*}
\begin{centering}
\includegraphics[scale=0.025]{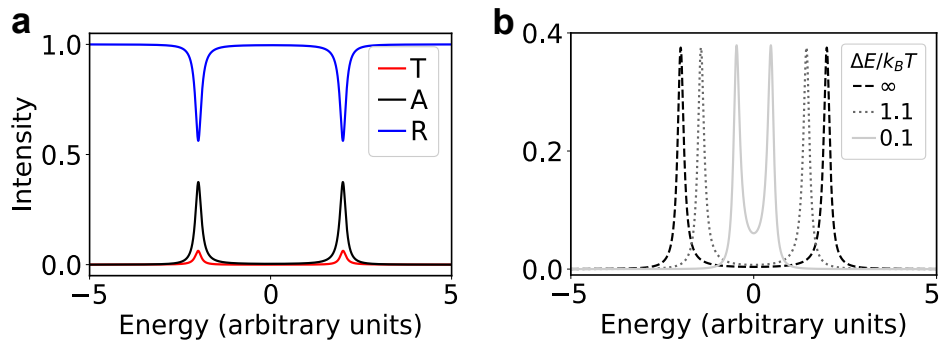}
\par\end{centering}
\caption{(a) Transmission, reflection, and absorption for an ensemble of identical
two-level systems, calculated with Eq. \ref{eq:T,A}. (b) Absorption
for an ensemble of identical two-level systems at different temperatures
demonstrating Rabi splitting contraction. The spectra have been computed
for resonant light and matter energies $\omega_{ph}=\omega_{exc}=0$,
cavity and molecule decay rates $\kappa=0.1$, $\gamma=0.3$, and
collective light-matter coupling $\sqrt{N}g=2$ (arbitrary frequency
units). \label{fig:Transmission,-reflection,-and}}
\end{figure*}
\begin{figure*}
\begin{centering}
\includegraphics[scale=0.2]{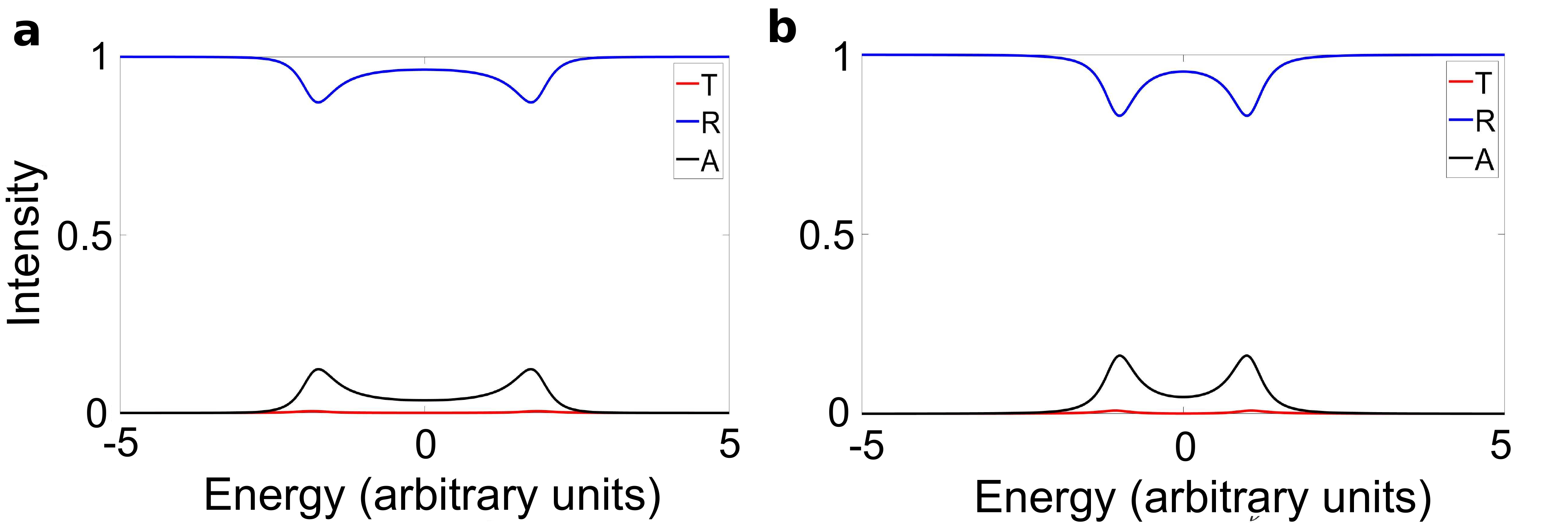}
\par\end{centering}
\caption{Transmission, reflection, and absorption for an ensemble of (a) Gaussian
energy-distributed two-level systems ($\sigma=1$, see Eq. \ref{eq:Gaussian})
and (b) Lorentzian energy-distributed two-level systems ($\sigma=1$,
see Eq. \ref{eq:Lorentzian}) obtained using Eqs. \ref{eq:spectroscopic_observables_harmonic-main}.
We show the examples with resonant light and matter energies $\omega_{ph}=\omega_{exc}=0$,
cavity and molecule decay rates $\kappa=0.1=0.1$, and collective
light-matter coupling $\sqrt{N}g=1.5$ (arbitrary frequency units).
\label{fig:TLS_disorder}}
\end{figure*}

As an follow-up to this model, let us now consider the effects of
a probability distribution $p(\omega_{exc})$ on the excitation energies
$\omega_{exc}$ at $T=0$; then,

\begin{equation}
\chi(\omega)=-N|g|^{2}\int d\omega_{exc}\frac{p(\omega_{exc})}{\omega-\omega_{exc}+i\frac{\gamma}{2}}.\label{eq:Spectral_function_disorder}
\end{equation}
We consider two types of disorder: Gaussian (Fig. \ref{fig:TLS_disorder}
a),

\begin{equation}
p(\omega_{exc})=\frac{1}{\sqrt{2\pi\sigma}}e^{-(\omega_{exc}-\bar{\omega})^{2}/2\sigma^{2}}\label{eq:Gaussian}
\end{equation}

and Lorentzian (Fig. \ref{fig:TLS_disorder} b),

\begin{equation}
p(\omega_{exc})=\frac{1}{\pi}\frac{(\sigma/2)^{2}}{(\omega_{exc}-\bar{\omega})^{2}+(\sigma/2)^{2}}.\label{eq:Lorentzian}
\end{equation}
Further aspects of disorder in cavities have been discussed in the
seminal work by \citep{houdre1996vacuum} and recently revisited in
many studies \citep{Cwik2016,Botzung2020,sommer2021molecular,du2022catalysis,engelhardt2022unusual,gera2022effects,gera2022exact,dubail2022large,cohn2022vibrational,chen2022interplay}. 

\subsection{$N\gg1$ Two-level molecules}

We generalize the Tavis-Cummings above and add one vibration coupled
to each electronic transition,

\begin{align*}
H_{mol} & =\sum_{i=1}^{N}\hbar\omega_{exc,i}\sigma_{i}^{\dagger}\sigma_{i}+\hbar\omega_{v}[b_{i}^{\dagger}b_{i}-\sqrt{S}\sigma_{i}^{\dagger}\sigma_{i}(b_{i}^{\dagger}+b_{i})]\\
V & =-\hbar\lambda(a+a^{\dagger})\sum_{i=1}^{N}\mu_{i}(\sigma_{i}^{\dagger}+\sigma_{i})+\text{h.c.}
\end{align*}
Here, $b_{i}^{\dagger}$ ($b_{i}$) is the creation (annhilation)
operator for a vibrational excitation in a high-frequency $\omega_{v}$
harmonic mode of the $i$-th molecule. The electron-vibration coupling
is characterized by the so-called Huang-Rhys parameter $S$. This
system has been previously studied by \citep{zeb2018exact} (see their
Fig. 8). To get started, we consider a $T=0$ ensemble of $N$ identical
molecules, $\omega_{exc,i}=E_{exc}$, $\mu_{i}=\mu$. Then, defining
$g^{2}=|\lambda\mu|^{2}$ the susceptibility is,

\begin{equation}
\chi(\omega)=-\sum_{m}\frac{Ng^{2}|\langle m'|0\rangle|^{2}}{\omega-(\omega_{exc}-S\omega_{v}+m\omega_{v})+i\frac{\gamma}{2}}.\label{eq:Spectral_function_DHO}
\end{equation}
Here, $|\langle m'|0\rangle|^{2}=e^{-S}\frac{S^{m}}{m!}$ is the Franck-Condon
factor for the $|0\rangle\to|m'\rangle$ vibronic transition, and
the vertical transition is at frequency $\omega_{exc}$, which contains
$S\omega_{v}$ vibrational quanta. The spectra coincides with that
obtained using the Collective dynamics Using Truncated Equations (CUT-E)
method developed to address the quantum dynamics of ensembles of $N$
complex molecules coupled to a cavity, in the limit when $O(N^{-k})$
effects ($k\geq1$) are ignored \citep{perez2023simulating}. 

\begin{figure}
\begin{centering}
\includegraphics[scale=0.09]{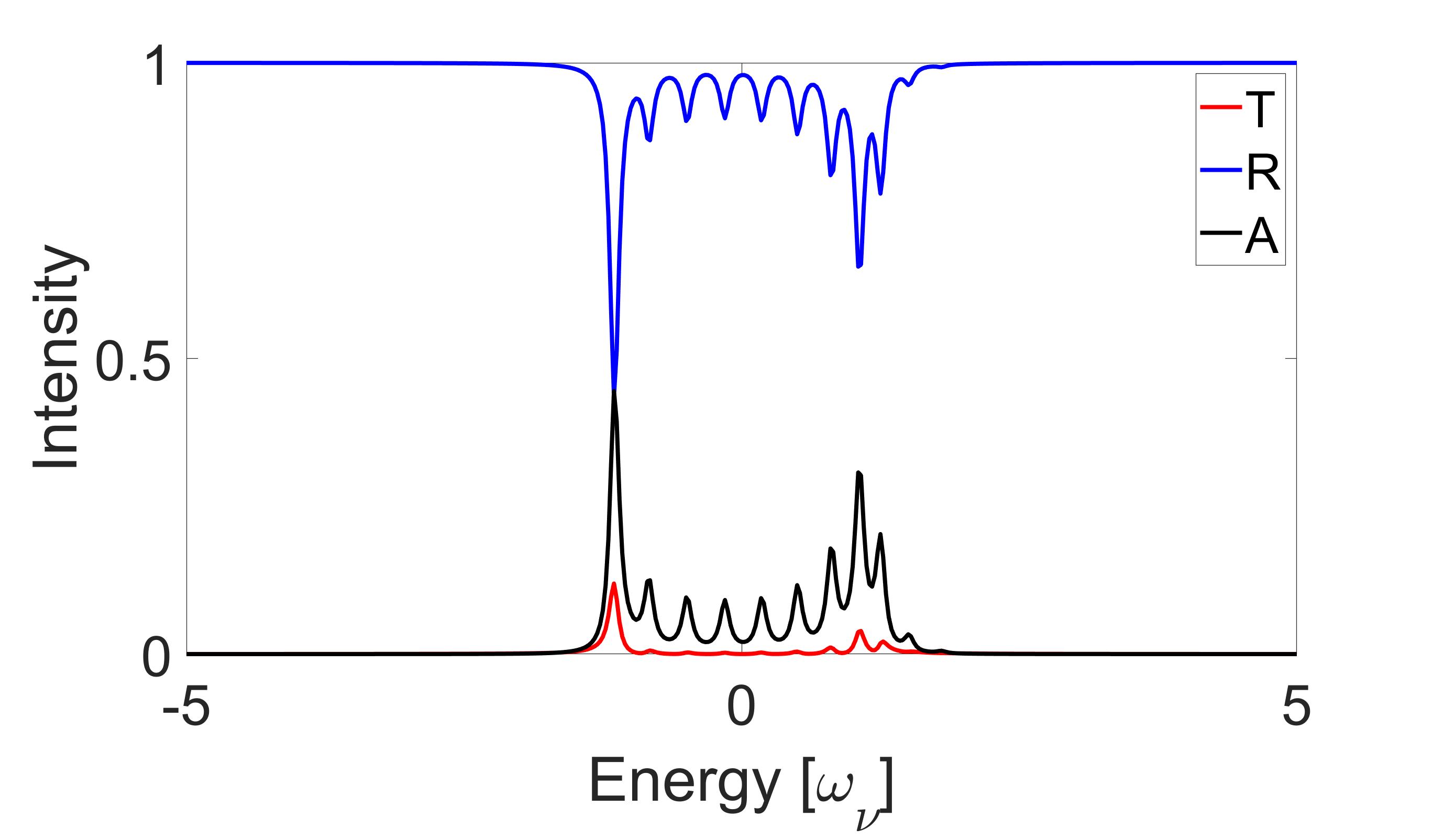}
\par\end{centering}
\caption{Transmission, reflection, and absorption for an ensemble of identical
two-level systems with vibronic coupling obtained using Eq. \ref{eq:spectroscopic_observables_harmonic-main}.
We show an example with resonant light and matter energies $\omega_{ph}=\omega_{exc}=0$,
cavity and molecule decay rates $\kappa=\gamma=0.1$, collective light-matter
coupling $\sqrt{N}g=1$ (arbitrary frequency units), $\omega_{v}=0.3$,
and $S=3$.}
\end{figure}

\subsection{$N\gg1$ Three-level systems}

\begin{figure*}
\begin{centering}
\includegraphics[scale=0.2]{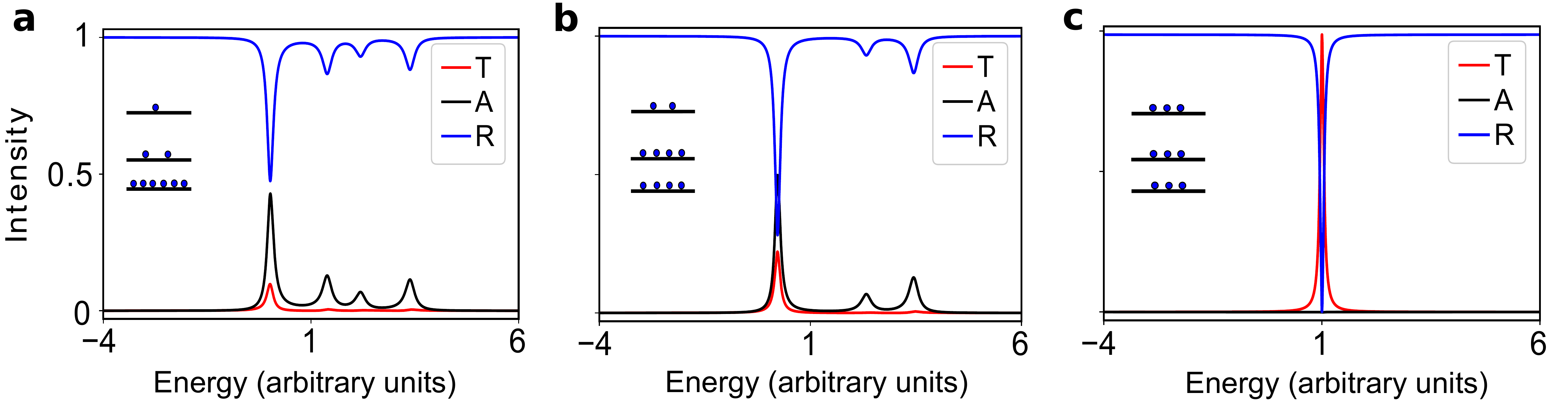}
\par\end{centering}
\caption{Transmission, reflection, and absorption for an ensemble of identical
three-level systems for (a) $p_{1}=0.7,$ $p_{2}=0.2,$ $p_{3}=0.1$,
(b) $p_{1}=0.48,$ $p_{2}=0.48,$ $p_{3}=0.04$, (c) $p_{1}=p_{2}=p_{3}$,
obtained using Eq. \ref{eq:spectroscopic_observables_harmonic-main}.
The spectra have been computed for the parameters, $\omega_{ph}=\omega_{12}=1$,
$\omega_{23}=2\omega_{12}$, cavity and molecule decay rates $\kappa=0.1$,
$\gamma=0.3$, collective light-matter coupling $\sqrt{N}g=1$ (arbitrary
frequency units).\label{fig:three_LS}}
\end{figure*}

We conclude our examples by considering an illustrative ensemble of
$N$ three-level systems placed in a cavity. The three optical transitions
couple to the cavity-photon mode. The Hamiltonian is given as \begin{subequations}\label{eq:three_level_system_H_V}

\begin{align}
H_{mol} & =\sum_{y=1}^{3}\sum_{i=1}^{N}\hbar\omega_{y,i}|y_{i}\rangle\langle y_{i}|,\label{eq:H_three_LS}\\
V & =-\hbar\lambda a\sum_{y,z=1,y\neq z}^{3}\sum_{i=1}^{N}\mu_{zy,i}|z_{i}\rangle\langle y_{i}|,\label{eq:V_three_LS}
\end{align}
\end{subequations}\noindent  where $\mu_{zy,i}=\langle z_{i}|\mu|y_{i}\rangle$
is the amplitude for the $|y\rangle\rightarrow|z\rangle$ transition
in the $i^{\text{th}}$ molecule. When all the the molecules in the
ensemble are identical, $\omega_{zy,i}=\omega_{zy}$ and $\mu_{zy,i}=\mu_{zy}$
for all $i$, the susceptibility $\chi(\omega)$ in Eq. \ref{eq:chi_2}
reads

\begin{equation}
\chi(\omega)=-\sum_{y,z=1}^{3}(p_{y}-p_{z})\frac{Ng^{2}|\mu_{zy}|^{2}}{\omega-\omega_{zy}+i\frac{\gamma}{2}}.\label{eq:chi_three_LS_identical}
\end{equation}
The transmission, absorption and reflection spectra of the setup have
been shown in Fig.\ref{fig:three_LS} for different population ratios.
For $p_{1}>p_{2}>p_{3}$, we see the four polariton peaks owing to
the three optical transitions at distinct frequencies coupling to
the cavity (Fig.\ref{fig:three_LS} a). Optical saturation of one
of the transitions ($p_{y}=p_{z}$) causes those transitions to become
transparent to the cavity (the inverse temperature for this transition
becomes $\beta_{yz}^{\text{eff}}=0$) (see Eq. \ref{eq:effective_temperature}
and the discussion about Rabi splitting contraction in Subsection
\ref{subsec:-Two-level-systems}), and show three polariton peaks
in the spectra (Fig. \ref{fig:three_LS}b). For $p_{1}=p_{2}=p_{3}$
all transitions are saturated, and the system behaves like an empty
cavity (Fig. \ref{fig:three_LS}c). If $p_{i}$ do not obey a Boltzmann
distribution, these states can in principle be obtained in optical
pumping experiments after dephasing decouples light and matter; if
they do (as in the last case), they can be also be obtained by tuning
the surroundings at the corresponding temperature $\mathcal{T}$.

\section{Conclusion\label{sec:Conclusion}}

In this article, we have treated the polariton problem as a quantum
impurity model where the photon is the impurity coupled to the optical
transitions of $N$ molecules. In the large $N$ limit, we have shown
that the photon Green's function can be trivially obtained with the
linear susceptibility $\chi(\omega)$ of the bare molecule (which
in turn, owing to Kramers Kronig relations, can be obtained solely
from bare molecular absorption spectra), bypassing a costly simulation
of $N$ explicit molecules coupled to a cavity (Eq. \ref{eq:spectroscopic_observables_harmonic-main}).
This result is quite general and is consistent with the success with
which transfer matrix methods in classical optics are used to model
polariton spectra. Importantly, it holds for arbitrary initial states
so long as the light and the matter are decoupled and stationary,
providing us with a very simple tool to understand a wide scope of
phenomena (Section \ref{sec:Examples}), ranging from Rabi splitting
features in idealized ensembles, but also complex lineshapes involving
optical saturation, disorder, vibronic coupling, and nonequilibrium
stationary states such as those that emerge upon optical pumping (\emph{i.e.},
the many situations where polaritons have relaxed to incoherent “dark
states,” yet they have not fully thermalized, as in ultrafast experiments).

For small $N$, $\chi(\omega)$ as the molecular information is not
enough to compute a polariton spectrum. However, the latter can still
be expressed in terms of a photon Green's function, as per Kubo linear
response formalism (Eq. \ref{eq:final_formulas-main}). These expressions
have been featured in previous studies \citep{Cwik2016,zeb2018exact},
but their derivation was not explicitly provided. We do so in the
Appendix and discuss their simplifications in the harmonic ($N\to\infty$)
regime. 

Finally, the fact that polaritons, regarded as a quantum impurity
problems, are simple when $N\to\infty$, is a possibly counterintuitive
yet intringuing observation \citep{gull2011continuous,segal2011nonequilibrium,cohen2011memory,gull2011numerically,hsieh2018unified,lindoy2018simple}.
In fact, recent work explores this fact to develop a mean-field approach
to polariton dynamics \citep{fowler2022efficient}. In another work,
we have recently shown that there is a hierarchy of timescales that
allows for the efficient simulation of molecular polaritons when $N\gg1$
but not infinite (Collective dynamics Using Truncated Equations, CUT-E
\citep{perez2023simulating,perezsanchez2023frequencydependent}),
giving rise to $O(N^{-k})$ rates that account for the finite size
of the molecular ensemble, very much in the spirit of $1/N$ expansions
in other fields including quantum field theory and quantum chemistry.
Understanding the class of quantum dynamics problems that afford similar
strategies is a fascinating direction to be explored in the near future.

\section{Acknowledgements}

This work was supported as part of the Center for Molecular Quantum
Transduction (CMQT), an Energy Frontier Research Center funded by
the U.S. Department of Energy, Office of Science, Basic Energy Sciences
under Award No. DE-SC0021314. We acknowledge key early discussions
with Nancy Makri where the connection between Ref. \citep{Makri1999}
and the molecular polariton problem was established. J.Y.Z. acknowledges
helpful discussions with Daniel Finkelstein Shapiro and Ignacio Franco
throughout the writing of the manuscript. A.K. thanks Kai Schwenickke
and Sindhana Pannir-Sivajothi for useful discussions. 

\section{Appendix: Derivation of spectroscopic observables \label{sec:Formulas-for-linear}}

We will assume the RWA throughout this Appendix.

\subsection{Input-output theory}

To keep this manuscript self-contained, we derive formulas for linear
spectroscopy of cavity polaritons using input-output (IO) theory \citep{Gardiner1985,Ciuti2006,portolan2008nonequilibrium,Li2018,SteckBook}.
We couple the molecular microcavity (Eq. \ref{eq:H-1}) with left
and right radiative continua via the photon mode,

\begin{equation}
H_{\text{total}}=H+H_{L}+H_{R},\label{eq:H'}
\end{equation}
where 

\begin{align}
H_{K}= & \int_{0}^{\infty}d\omega'\hbar\omega'b^{K\dagger}(\omega')b^{K}(\omega')\nonumber \\
 & +\Bigg[\hbar\frac{\sqrt{\kappa_{K}}}{\sqrt{2\pi}}\int_{0}^{\infty}d\omega'b^{K\dagger}(\omega')a+\text{h.c.}\Bigg]\label{eq:HK}
\end{align}
for $K=L,R$ only feature RWA terms and 

\begin{subequations}\label{eq:commutation}
\begin{align}
[b^{K}(\omega),b^{K'}(\omega')] & =[b^{K\dagger}(\omega),b^{K'\dagger}(\omega')]=0,\label{eq:commutation1}\\{}
[b^{K}(\omega),b^{K'\dagger}(\omega')] & =\delta_{KK'}\delta(\omega-\omega').\label{eq:commutation_2}
\end{align}
\end{subequations}\noindent For convenience, we will now derive some
results in the Heisenberg picture (corresponding to evolution with
respect to $H_{\text{total}})$, with the corresponding operators
labeled by the subscript H, \emph{e.g.}, $a_{\text{H}}(t)=e^{-iH_{\text{total}}(t-t_{\text{in}})/\hbar}ae^{iH_{\text{total}}(t-t_{\text{in}})/\hbar}$.
Schrödinger picture operators will continue to be indicated without
an explicit subscript. The Equation of Motion (EoM) for the cavity
photon is

\begin{align}
\frac{\partial a_{\text{H}}(t)}{\partial t}= & -\frac{i}{\hbar}[a_{\text{H}}(t),H_{\text{H}}(t)]\nonumber \\
 & -i\sum_{K=L,R}\sqrt{\frac{\kappa_{K}}{2\pi}}\int_{0}^{\infty}d\omega'b_{\text{H}}^{K}(\omega')(t),\label{eq:photon_EoM}
\end{align}
Similarly, the corresponding EoM for the bath mode $b_{\text{H}}^{K}(\omega)(t)$
is

\begin{align}
\frac{\partial b_{\text{H}}^{K}(\omega)(t)}{\partial t} & =-i\omega b_{\text{H}}^{K}(\omega)-i\sqrt{\frac{\kappa_{K}}{2\pi}}a_{\text{H}}(t)\nonumber \\
\implies\frac{\partial[b_{\text{H}}^{K}(\omega)(t)e^{i\omega t}]}{\partial t} & =-i\sqrt{\frac{\kappa_{K}}{2\pi}}a_{\text{H}}(t)e^{i\omega t}.\label{eq:bath mode}
\end{align}
Defining $t_{\text{in}}<t$ and $t_{\text{out}}>t$, we can integrate
Eq. \ref{eq:bath mode} to obtain,

\begin{subequations}\label{eq:bath_integrated}

\begin{align}
b_{\text{H}}^{K}(\omega)(t)= & b_{\text{H}}^{K}(\omega)(t_{\text{in}})e^{-i\omega(t-t_{\text{in}})}\nonumber \\
 & -i\sqrt{\frac{\kappa_{K}}{2\pi}}\int_{t_{\text{in}}}^{t}dt'a_{\text{H}}e^{-i\omega(t-t')},\label{eq:bath_in}\\
b_{\text{H}}^{K}(\omega)(t)= & b_{\text{H}}^{K}(\omega)(t_{\text{out}})e^{-i\omega(t-t_{\text{out}})}\nonumber \\
 & +i\sqrt{\frac{\kappa_{K}}{2\pi}}\int_{t}^{t_{\text{out}}}dt'a_{\text{H}}e^{-i\omega(t-t')}.\label{eq:bath_out}
\end{align}
\end{subequations}\noindent Let us feed Eq. \ref{eq:bath_integrated}
for $K=L$ into Eq. \ref{eq:photon_EoM}. By approximating $\int_{0}^{\infty}d\omega'\approx\int_{-\infty}^{\infty}d\omega'$,

\begin{subequations}\label{eq:a_1_a_2}

\begin{align}
\frac{\partial a_{\text{H}}(t)}{\partial t}= & -\frac{i}{\hbar}[a_{\text{H}}(t),H_{\text{H}}(t)]-i\sqrt{\frac{\kappa_{R}}{2\pi}}\int_{0}^{\infty}d\omega'b_{\text{H}}^{R}(\omega')(t)\nonumber \\
 & -\frac{\kappa_{L}}{2}a_{\text{H}}(t)-\sqrt{\kappa_{L}}b_{\text{in,H}}^{L}(t)\label{eq:a_1}\\
= & -\frac{i}{\hbar}[a_{\text{H}}(t),H_{\text{H}}(t)]-i\sqrt{\frac{\kappa_{R}}{2\pi}}\int_{0}^{\infty}d\omega'b_{\text{H}}^{R}(\omega')(t)\nonumber \\
 & +\frac{\kappa_{L}}{2}a_{\text{H}}(t)-\sqrt{\kappa_{L}}b_{\text{out,H}}^{L}(t).\label{eq:a_2}
\end{align}
\end{subequations}\noindent where

\begin{subequations}\label{eq:IO_operators}

\begin{align}
b_{\text{in,H}}^{K}(t) & =\frac{i}{\sqrt{2\pi}}\int_{-\infty}^{\infty}d\omega'b_{\text{H}}^{K}(\omega)(t_{\text{in}})e^{-i\omega(t-t_{\text{in}})},\label{eq:Input}\\
b_{\text{out,H}}^{K}(t) & =\frac{i}{\sqrt{2\pi}}\int_{-\infty}^{\infty}d\omega'b_{\text{H}}^{K}(\omega)(t_{\text{out}})e^{-i\omega(t-t_{\text{out}})}.\label{eq:Output}
\end{align}
\end{subequations}\noindent By comparing Eqs. \ref{eq:a_1} and \ref{eq:a_2},
we obtain the IO relations for the left continuum,

\begin{subequations}\label{eq:IO_relations}

\begin{equation}
b_{\text{out,H}}^{L}(t)-b_{\text{in,H}}^{L}(t)=\sqrt{\kappa_{L}}a_{\text{H}}(t).\label{eq:IO_left}
\end{equation}
It is clear that repeating the procedure of Eqs. \ref{eq:a_1_a_2}–\ref{eq:IO_operators}
with $K=R$ yields the analogous IO relations for the right continuum,

\begin{equation}
b_{\text{out,H}}^{R}(t)-b_{\text{in,H}}^{R}(t)=\sqrt{\kappa_{R}}a_{\text{H}}(t).\label{eq:IO_right}
\end{equation}

\end{subequations}\noindent For completeness, we can also express
Eq. \ref{eq:photon_EoM} in terms of the IO operators of both continua,

\begin{subequations}\label{eq:EoM_a_IO}

\begin{align}
\frac{\partial a_{\text{H}}(t)}{\partial t}= & -\frac{i}{\hbar}[a_{\text{H}}(t),H_{\text{H}}(t)]+\sum_{K}\Bigg[-\frac{\kappa_{K}}{2}a_{\text{H}}(t)-\sqrt{\kappa_{K}}b_{\text{in,H}}^{K}(t)\Bigg]\label{eq:EoM_a_input}\\
= & -\frac{i}{\hbar}[a_{\text{H}}(t),H_{\text{H}}(t)]+\sum_{K}\Bigg[\frac{\kappa_{K}}{2}a_{\text{H}}(t)-\sqrt{\kappa_{K}}b_{\text{out,H}}^{K}(t)\Bigg].\label{eq:EoM_a_output}
\end{align}
\end{subequations}\noindent Assuming that the density matrix at $t=t_{\text{in}}$
is a product state between the molecular microcavity and the continua,

\begin{equation}
\rho_{\text{total}}(t_{\text{in}})=\rho_{L}(t_{\text{in}})\otimes\rho(t_{\text{in}})\otimes\rho_{R}(t_{\text{in}}),\label{eq:initial wavefunction}
\end{equation}
and that the driving occurs only from the left continuum, 

\begin{subequations}\label{eq:b_initial_conditions}
\begin{align}
\langle b_{\text{in,H}}^{L}(t)\rangle & \neq0,\label{eq:b_in_L_0}\\
\langle b_{\text{in,H}}^{R}(t)\rangle & =0,\label{eq:b_out_L_0}
\end{align}

\end{subequations}\noindent we get, after tracing over the continua,

\begin{align}
\frac{\partial a_{\text{H}}(t)}{\partial t}= & -\frac{i}{\hbar}[a_{\text{H}}(t),H_{\text{H}}(t)]-\frac{\kappa}{2}a_{\text{H}}(t)-\sqrt{\kappa_{L}}\langle b_{\text{in,H}}^{L}(t)\rangle,\label{eq:da/dt_Langevin}
\end{align}
where $\kappa=\kappa_{L}+\kappa_{R}$. Eq. \ref{eq:da/dt_Langevin}
can be rewritten as,
\begin{align}
\frac{\partial a_{\text{H}}(t)}{\partial t}= & -\frac{i}{\hbar}[a_{\text{H}}(t),\tilde{H}_{\text{H}}(t)],\label{eq:a_equation_EFFECTIVE}
\end{align}
which allows us to conclude that the effective time-dependent Hamiltonian
(in the Schrödinger picture) governing the molecular microcavity is
\begin{subequations}\label{eq:H_tilde_eqs}
\begin{equation}
\tilde{H}(t)=H'+H_{int}(t).\label{eq:H_tilde}
\end{equation}
In the absence of drive, the molecular microcavity obeys the effective
non-Hermitian Hamiltonian
\begin{equation}
H'=H-i\frac{\hbar\kappa}{2}a^{\dagger}a,\label{eq:non-Hermitian}
\end{equation}
while the time-dependent drive of the cavity due to light coupling
from the left hand side is 
\begin{equation}
H_{int}(t)=-i\hbar\sqrt{\kappa_{L}}\langle b_{\text{in,H}}^{L}(t)\rangle a^{\dagger}+\text{h.c.}\label{eq:V(t)}
\end{equation}
\end{subequations}\noindent Equipped with this formalism, we are
interested in computing the following spectroscopic observables,

\begin{subequations}\label{eq:spectroscopic_observables}

\begin{align}
T(\omega) & =\frac{|\langle b_{\text{out,H}}^{R}(\omega)\rangle|^{2}}{|\langle b_{\text{in,H}}^{L}(\omega)\rangle|^{2}}\nonumber \\
 & =\frac{\kappa_{R}|\langle a_{\text{H}}(\omega)\rangle|^{2}}{|\langle b_{\text{in,H}}^{L}(\omega)\rangle|^{2}},\label{eq:T(w)}\\
R(\omega) & =\frac{|\langle b_{\text{out,H}}^{L}(\omega)\rangle|^{2}}{|\langle b_{\text{in,H}}^{L}(\omega)\rangle|^{2}}\nonumber \\
 & =\frac{|\langle\sqrt{\kappa_{L}}a_{\text{H}}(\omega)+b_{\text{in,H}}^{L}(\omega)\rangle|^{2}}{|\langle b_{\text{in,H}}^{L}(\omega)\rangle|^{2}},\label{eq:R(w)}\\
A(\omega) & =1-T(\omega)-R(\omega),\label{eq:A(w)}
\end{align}
\end{subequations}\noindent where the traces above are carried out
with respect to the initial state (Eq. \ref{eq:initial wavefunction}),
$\langle\cdot\rangle=\text{Tr}[\cdot\rho_{\text{total}}(t_{\text{in}})]$,
and in particular, $\langle a_{\text{H}}(\omega)\rangle=\text{Tr}[a_{\text{H}}(\omega)\rho(t_{\text{in}})]$
only depends on the initial state of the molecular microcavity. We
have also used Eq. \ref{eq:b_initial_conditions} and the Fourier
transform convention in Eq. \ref{eq:FT}. Eq. \ref{eq:spectroscopic_observables}
reveals that all the relevant spectroscopic observables can be obtained
once $\langle a_{\text{H}}(\omega)\rangle$ is known. We now show
two scenarios where $\langle a_{\text{H}}(\omega)\rangle$ can be
easily computed.

\subsection{$\langle a_{\text{H}}(\omega)\rangle$ from Kubo linear response}

\subsubsection{Derivation of transmission, reflection, and absorption formulas}

Hereafter we set $t_{\text{in}}=0$. In the general case, $H$ contains
anharmonic terms, so the evaluation of $\langle a_{\text{H}}(\omega)\rangle$
cannot be performed exactly. Instead, we can carry out a perturbation
expansion in $H_{int}$ for each of the Heisenberg operators in Eq.
\ref{eq:a_equation_EFFECTIVE} and solve for $a_{\text{H}}^{(n)}(t)$
up to lowest nonvanishing order $n$. 

Starting at zeroth-order $O(H_{int}^{0})$,

\begin{align}
\frac{\partial a_{\text{H}}^{(0)}(t)}{\partial t}-\frac{i}{\hbar}[H',a_{\text{H}}^{(0)}(t)] & =0\label{eq:zeroth_order_ODE}
\end{align}
can be solved by,

\begin{equation}
a_{\text{H}}^{(0)}(t)=e^{iH't/\hbar}ae^{-iH't/\hbar}.\label{eq:zeroth_order_solution}
\end{equation}
Recall our assumption that the initial molecular microcavity state
is a product state between photon and molecular degrees of freedom
(see \ref{eq:product_state_section-1}), $\rho(0)=\rho_{ph}\otimes\rho_{mol}$,

\begin{equation}
e^{-iH't/\hbar}\rho(0)\stackrel{t\to\infty}{=}|0\rangle\langle\varphi_{\text{ph}}|\otimes\rho_{mol}.\label{eq:no_photons!}
\end{equation}
where $|\varphi_{\text{ph}}\rangle$ is a photonic state. Then,

\begin{align}
\langle a_{\text{H}}^{(0)}(t)\rangle & =\text{Tr}[e^{iH't/\hbar}ae^{-iH't/\hbar}\rho(t_{\text{in}})]\nonumber \\
 & \stackrel{t\to\infty}{=}0,\label{eq:zeroth_order}
\end{align}
which makes sense since any transient photonic amplitude will vanish
due to photon escape.

Similarly, at $O(H_{int})$, we have

\begin{align}
\frac{\partial a_{\text{H}}^{(1)}(t)}{\partial t}-\frac{i}{\hbar}[H',a_{\text{H}}^{(1)}(t)] & =-\sqrt{\kappa_{L}}\langle b_{\text{in,H}}^{L}(t)\rangle.\label{eq:first_order_ODE}
\end{align}
This is a first order inhomogeneous differential equation that can
be solved with Green's function methods. Defining

\begin{equation}
G^{R}(t)=\Theta(t)[a_{\text{H}}^{(0)}(t),a_{\text{H}}^{(0)\dagger}(0)],\label{eq:DR(t)}
\end{equation}
which solves,

\begin{equation}
\frac{\partial G^{R}(t-t')}{\partial t}-\frac{i}{\hbar}[H',D^{R}(t-t')]=\delta(t-t').\label{eq:delta_impulse}
\end{equation}
where we need an additional assumption: the trace is performed over
an initial state $\rho(t_{\text{in}})$ that is stationary with respect
to $H'$; thus, it contains no photons. We readily obtain the Kubo
linear response formula,

\begin{align}
\langle a^{(1)}(t)\rangle= & -\sqrt{\kappa_{L}}\int_{-\infty}^{\infty}dt_{1}\langle b_{\text{in}}^{L}(t-t_{1})\rangle D^{R}(t_{1})\nonumber \\
= & -\sqrt{\kappa_{L}}\int_{-\infty}^{\infty}dt_{1}\langle b_{\text{in}}^{L}(t-t_{1})\rangle\Theta(t_{1})\langle[a_{\text{H}}^{(0)}(t_{1}),a^{\dagger}]\rangle.\label{eq:first_order_eq}
\end{align}
where $D^{R}(t)=\langle G^{R}(t)\rangle$ is the retarded Green's
function. Importantly, Eq. \ref{eq:first_order_eq} has the form of
a convolution, 

\begin{equation}
\langle a(\omega)\rangle=\langle a_{\text{H}}^{(1)}(\omega)\rangle=-i\sqrt{\kappa_{L}}\langle b_{\text{in,H}}^{L}(\omega)\rangle D^{R}(\omega),\label{eq:a_omega_perturbative}
\end{equation}
where $D^{R}(\omega)$, according to our Fourier transform (\ref{eq:FT})
convention, reads

\begin{align}
D^{R}(\omega) & =-i\int_{-\infty}^{\infty}dte^{i\omega t}\Theta(t)\langle[e^{iH't/\hbar}ae^{-iH't/\hbar},a^{\dagger}]\rangle.\label{eq:D_R(omega)}
\end{align}
Incidentally, given the non-Hermitian nature of $H'$ due to photon
leakage, $\rho(0)$ cannot contain photons, so one of the terms in
the commutator is superfluous and the final propagator can be replaced,
$H'\to H$, 

\begin{align}
D^{R}(\omega) & =-i\int_{-\infty}^{\infty}dte^{i\omega t}\Theta(t)\langle e^{iHt/\hbar}ae^{-iH't/\hbar}a^{\dagger}\rangle.\label{eq:D_R(omega)_simple}
\end{align}

Eq. \ref{eq:D_R(omega)_simple} can be fed into Eq. \ref{eq:a_omega_perturbative}.
Using Eq. \ref{eq:IO_relations}, \ref{eq:b_initial_conditions},
and \ref{eq:spectroscopic_observables} gives \citep{Cwik2016,zeb2018exact},
\begin{subequations}\label{eq:final_formulas}

\begin{align}
T(\omega) & =\kappa_{L}\kappa_{R}|D^{R}(\omega)|^{2},\label{eq:T_w}\\
R(\omega) & =1+2\kappa_{L}\Im D^{R}(\omega)+\kappa_{L}^{2}|D^{R}(\omega)|^{2},\label{eq:R_w}\\
A(\omega) & =-\kappa_{L}[\kappa|D^{R}(\omega)|^{2}+2\Im D^{R}(\omega)].\label{eq:A_w}
\end{align}

\end{subequations}\noindent  

\subsubsection{A “Landauer” formula}

We now re-express the photon retarded Green function (Eq. \ref{eq:D_R(omega)})
as,

\begin{align}
D^{R}(\omega) & =-i\int_{-\infty}^{\infty}dte^{i\omega t}\Theta(t)\langle[a(t),a^{\dagger}]\rangle\nonumber \\
 & =-i\int_{-\infty}^{\infty}dte^{i\omega t}\Theta(t)\langle\mathcal{A}(t)a^{\dagger}\rangle\nonumber \\
 & =-i\int_{-\infty}^{\infty}dte^{i\omega t}\Theta(t)\langle e^{i\mathcal{L}'t/\hbar}\mathcal{A}e^{-i\mathcal{L}'t/\hbar}a^{\dagger}\rangle\nonumber \\
 & =\langle\mathcal{A}\mathcal{G}(\omega)a^{\dagger}\rangle,\label{eq:D_R_as_Geff}
\end{align}
where we have introduced Liouville space operators $\mathcal{A}(t)=[a(t),\cdot]$
and $\mathcal{L}'=[H',\cdot]$ (see \citep{MukamelBook} Chapter 3).
We also assumed stationarity of $\rho(0)$ under evolution with respect
to $\mathcal{L}'$. Furthermore, we introduced $\mathcal{G}(\omega)$
as the frequency domain retarded Green function

\begin{equation}
\mathcal{G}(\omega)=\frac{1}{\omega-\mathcal{L}'/\hbar}.\label{eq:Liouville G}
\end{equation}
Plugging Eq. \ref{eq:D_R_as_Geff} into Eq. \ref{eq:T(w)}, we obtain
a Liouville-space “Landauer” formula for transmission (see for
instance, Appendix 9B in \citep{NitzanBook}),

\begin{align}
T(\omega) & =\kappa_{L}\kappa_{R}|D^{R}(\omega)|^{2}\nonumber \\
 & =\kappa_{L}\kappa_{R}|\langle\mathcal{A}\mathcal{G}(\omega)a^{\dagger}\rangle|^{2}.\label{eq:Landauer_Liouville}
\end{align}

\subsection{$\langle a_{\text{H}}(\omega)\rangle$ from harmonic degrees of freedom
\label{subsec:effective harmonic}}

As argued in Section \ref{sec:Molecular-polaritons-as}, when the
number of molecules $N\to\infty$, the anharmonic molecular degrees
of freedom can be replaced by an effective harmonic bath. Then, the
effective molecular microcavity Hamiltonian with photon loss and RWA
reads (Eqs. \ref{eq:H_eff}–\ref{eq:Veff}),

\begin{align}
H^{\text{eff}\prime}= & \ensuremath{\hbar(\omega_{ph}-i\frac{\kappa}{2})}a^{\dagger}a+\sum_{j}\hbar\omega_{j}b_{j}^{\dagger}b_{j}\nonumber \\
 & -\Bigg[a\sum_{j}\hbar\bar{c}_{j}b_{j}^{\dagger}+\text{h.c.}\Bigg].\label{eq:H_harmonic_with_Veff}
\end{align}
The conclusions from this part \ref{subsec:effective harmonic} of
the Appendix will clearly also hold for idealized harmonic Hamiltonians
$H'$, where the $N\to\infty$ restriction is not needed.

\subsubsection{$\langle a_{\text{H}}(\omega)\rangle$ from EoM \label{subsub EoM}}

$\langle a_{\text{H}}(\omega)\rangle$ can be solved exactly for Eq.
\ref{eq:H_harmonic_with_Veff}. Using it to evaluate the $[a,H^{\text{eff}}]$
commutator in Eq. \ref{eq:da/dt_Langevin} yields,

\begin{subequations}\label{eq:aH_harmonic_EOM}

\begin{align}
\frac{\partial a_{\text{H}}(t)}{\partial t} & =-i(\omega_{ph}-i\frac{\kappa}{2})a_{\text{H}}(t)\nonumber \\
 & +i\bar{c}_{j}^{*}b_{j,\text{H}}-\sqrt{\kappa_{L}}b_{\text{in,H}}^{L}(t)\label{eq:aH(t)}\\
\implies-i\omega'a_{\text{H}}(\omega) & =-i(\omega_{ph}-i\frac{\kappa}{2})a_{\text{H}}(\omega)\nonumber \\
 & +\sum_{j}i\bar{c}_{j}^{*}b_{j,\text{H}}(\omega)-\sqrt{\kappa_{L}}b_{\text{in,H}}^{L}(\omega).\label{eq:aH(omega)}
\end{align}
\end{subequations}\noindent The corresponding EoM for each of the
$b_{j,\text{H}}(t)$ is given by,

\begin{subequations}\label{eq:bjH_harmonic_EOM}

\begin{align}
\frac{\partial b_{j,\text{H}}(t)}{\partial t} & =-i(\omega_{j}-i\frac{\gamma}{2})b_{j,\text{H}}(t)+i\bar{c}_{j}a_{\text{H}}(t)\label{eq:bjH(t)}\\
\implies-i\omega b_{j,\text{H}}(\omega) & =-i(\omega_{j}-i\frac{\gamma}{2})b_{j,\text{H}}(\omega)+i\bar{c}_{j}a_{\text{H}}(\omega).\label{eq:bjH(omega)}
\end{align}
\end{subequations}\noindent Solving Eq. \ref{eq:bjH(omega)} for
$b_{j,\text{H}}(\omega)$ and plugging the result into Eq. \ref{eq:aH(omega)}
yields the result,

\begin{align}
\langle a_{\text{H}}(\omega)\rangle= & \frac{-i\sqrt{\kappa_{L}}\langle b_{\text{in,H}}^{L}(\omega)\rangle}{(\omega-\omega_{c}+i\frac{\kappa}{2}-\Sigma_{M})}\label{eq:a_H(omega)_harmonic}
\end{align}
where the molecular self-energy is shown to be minus the linear susceptibility
of the molecules,

\begin{align}
\Sigma_{M} & =\sum_{j}\frac{|\bar{c}_{j}|^{2}}{\omega'-\omega_{j}+i\frac{\gamma}{2}}\nonumber \\
 & =-\chi(\omega).\label{eq:self_energy_as_minus_susceptibility}
\end{align}
As expected for a harmonic system, its response to driving is linear,
in this case, proportional to $\sqrt{\frac{\kappa}{2}}\langle b_{\text{in,H}}^{L}(\omega)\rangle$.
Interestingly, $\langle a_{\text{H}}(\omega)\rangle$ is independent
of the initial quantum state of the effective harmonic oscillators.
However, recall that the information about the initial thermal state
of the real anharmonic degrees of freedom is hidden in $\{\bar{c}_{j}\}$.

\subsubsection{$\langle a_{\text{H}}(\omega)\rangle$ from Kubo formula \label{subsub:-from-Kubo}}

We now show that the Kubo formula in Eq. \ref{eq:a_omega_perturbative},
despite being derived under a perturbation theory and initial stationary
state $\rho(t_{\text{in}})$, is not an approximation, but rather
gives the exact response if the molecular degrees of freedom can be
treated as harmonic. We invoke the analogous identity of Eq. \ref{eq:G=00003DG0+...}
but in Liouville space (to unclutter notation, we drop the $\omega$
argument hereafter when it is clear) to expand Eq. \ref{eq:D_R_as_Geff}, 

\begin{align}
\mathcal{G} & =\mathcal{G}_{0}+\mathcal{G}_{0}\frac{\mathcal{V}}{\hbar}\mathcal{G}_{0}+\mathcal{G}_{0}\frac{\mathcal{V}}{\hbar}\mathcal{G}_{0}\frac{\mathcal{V}}{\hbar}\mathcal{G},\label{eq:Liouville_G=00003DG0+...}
\end{align}
where the non-interacting Green function $\mathcal{G}_{0}$ corresponds
to \emph{$\mathcal{L}_{0}=[H_{0}',\cdot]=[H_{0}-i\hbar\frac{\kappa}{2}a^{\dagger}a,\cdot]$}.
This identity implies,

\begin{align}
\mathcal{A}\mathcal{G}a^{\dagger} & =\mathcal{A}\mathcal{G}_{0}a^{\dagger}+\mathcal{A}\mathcal{G}_{0}\frac{\mathcal{V}}{\hbar}\mathcal{G}_{0}a^{\dagger}+\mathcal{A}\mathcal{G}_{0}\frac{\mathcal{V}}{\hbar}\mathcal{G}_{0}\frac{\mathcal{V}}{\hbar}\mathcal{G}a^{\dagger}.\label{eq:AGa+}
\end{align}
Terms involving $\mathcal{G}_{0}$ can be readily and explicitly evaluated
in Hilbert space,

\begin{align}
\mathcal{A}\mathcal{G}_{0}^{\text{eff}}a^{\dagger}= & -i\int_{-\infty}^{\infty}dte^{i\omega t}\Theta(t)[e^{iH_{0}'t/\hbar}ae^{-iH_{0}'t/\hbar},a^{\dagger}]\nonumber \\
= & -i\int_{-\infty}^{\infty}dte^{i\omega t}\Theta(t)[a,a^{\dagger}]e^{-i(\omega_{ph}'-i\kappa/2)t}\nonumber \\
= & \frac{1}{\omega-\omega_{ph}+i\frac{\kappa}{2}}.\label{eq:AG0a+}
\end{align}

\begin{align}
\mathcal{B}_{j}\mathcal{G}_{0}^{\text{eff}}b_{j'}^{\dagger}= & -i\int_{-\infty}^{\infty}dte^{i\omega t}\Theta(t)[e^{iH_{0}'t/\hbar}b_{j}e^{-iH_{0}'t/\hbar},b_{j'}^{\dagger}]\nonumber \\
= & -i\int_{-\infty}^{\infty}dte^{i\omega t}\Theta(t)[b_{j},b_{j'}^{\dagger}]e^{-i(\omega_{ph}'-i\gamma/2)t}\nonumber \\
= & \frac{\delta_{jj'}}{\omega-\omega_{j}+i\frac{\gamma}{2}}.\label{eq:BjG0bj+}
\end{align}
The second term in the sum of Eq. \ref{eq:AGa+} can be shown to vanish
identically given that $V^{\text{eff}}$ couples photon with matter
degrees of freedom (but not matter with matter or photon with photon),

\begin{align}
 & \mathcal{A}\mathcal{G}_{0}^{\text{eff}}\frac{\mathcal{V}^{\text{eff}}}{\hbar}\mathcal{G}_{0}^{\text{eff}}a^{\dagger}\nonumber \\
= & \mathcal{A}\mathcal{G}_{0}^{\text{eff}}(a^{\dagger}\mathcal{A})\frac{\mathcal{V}^{\text{eff}}}{\hbar}(a^{\dagger}\mathcal{A})\mathcal{G}_{0}^{\text{eff}}a^{\dagger}\nonumber \\
= & (\mathcal{A}\mathcal{G}_{0}^{\text{eff}}a^{\dagger})(\mathcal{A}\frac{\mathcal{V}^{\text{eff}}}{\hbar}a^{\dagger})(\mathcal{A}\mathcal{G}_{0}^{\text{eff}}a^{\dagger})\nonumber \\
= & \Big(\frac{1}{\omega-\omega_{ph}+i\frac{\kappa}{2}}\Big)^{2}\mathcal{A}\frac{\mathcal{V}^{\text{eff}}}{\hbar}a^{\dagger}\nonumber \\
= & \Big(\frac{1}{\omega-\omega_{ph}+i\frac{\kappa}{2}}\Big)^{2}\mathcal{A}[\frac{V^{\text{eff}}}{\hbar},a^{\dagger}]\nonumber \\
= & \Big(\frac{1}{\omega-\omega_{ph}+i\frac{\kappa}{2}}\Big)^{2}\Big[a,-\sum_{j}\bar{c}_{j}b_{j}^{\dagger}\Big]\nonumber \\
= & 0.\label{eq:AG0VG0a}
\end{align}
where we inserted two “minus” resolutions of the identity in
the second line,

\begin{subequations}\label{eq:res_identity}

\begin{align}
\mathcal{A}a^{\dagger} & =[a,a^{\dagger}]\nonumber \\
 & =1\label{eq:res_identity_1}\\
\implies a^{\dagger}\mathcal{A} & =(\mathcal{A}^{\dagger}a)^{\dagger}\nonumber \\
 & =[a^{\dagger},a]^{\dagger}\nonumber \\
 & =-1\label{eq:res_identity_2}
\end{align}
\end{subequations}\noindent Finally, the third term in the sum of
Eq. \ref{eq:AGa+} is non-zero and leads to the self-energy of the
photon due to its coupling to the matter degrees of freedom,

\begin{align}
 & \mathcal{A}\mathcal{G}_{0}^{\text{eff}}\frac{\mathcal{V}^{\text{eff}}}{\hbar}\mathcal{G}_{0}^{\text{eff}}\frac{\mathcal{V}^{\text{eff}}}{\hbar}\mathcal{G}^{\text{eff}}a^{\dagger}\nonumber \\
= & \mathcal{A}\mathcal{G}_{0}^{\text{eff}}(a^{\dagger}\mathcal{A})\frac{\mathcal{V}^{\text{eff}}}{\hbar}\mathcal{G}_{0}^{\text{eff}}\frac{\mathcal{V}^{\text{eff}}}{\hbar}(a^{\dagger}\mathcal{A})\mathcal{G}^{\text{eff}}a^{\dagger}\nonumber \\
= & (\mathcal{A}\mathcal{G}_{0}^{\text{eff}}a^{\dagger})\mathcal{A}\frac{\mathcal{V}^{\text{eff}}}{\hbar}\mathcal{G}_{0}^{\text{eff}}(\frac{\mathcal{V}^{\text{eff}}}{\hbar}a^{\dagger})(\mathcal{A}\mathcal{G}^{\text{eff}}a^{\dagger})\nonumber \\
= & -\sum_{j}(\mathcal{A}\mathcal{G}_{0}^{\text{eff}}a^{\dagger})(\mathcal{A}\frac{\mathcal{V}^{\text{eff}}}{\hbar}b_{j}^{\dagger})\mathcal{B}_{j}\mathcal{G}_{0}^{\text{eff}}(-\bar{c}_{j}b_{j}^{\dagger})(\mathcal{A}\mathcal{G}^{\text{eff}}a^{\dagger})\nonumber \\
= & -\sum_{j}(\mathcal{A}\mathcal{G}_{0}^{\text{eff}}a^{\dagger})\mathcal{A}(-\bar{c}_{j}a^{\dagger})\mathcal{B}_{j}\mathcal{G}_{0}^{\text{eff}}(-\bar{c}_{j}b_{j}^{\dagger})(\mathcal{A}\mathcal{G}^{\text{eff}}a^{\dagger})\nonumber \\
= & -\Big(\frac{1}{\omega-\omega_{ph}+i\frac{\kappa}{2}}\Big)\sum_{j}\Big(\frac{|\bar{c}_{j}|^{2}}{\omega-\omega_{j}+i\frac{\gamma}{2}}\Big)(\mathcal{A}\mathcal{G}^{\text{eff}}a^{\dagger}).\nonumber \\
= & -\Big(\frac{1}{\omega-\omega_{ph}+i\frac{\kappa}{2}}\Big)\Sigma_{M}(\mathcal{A}\mathcal{G}^{\text{eff}}a^{\dagger}).\label{eq:AGVGVGa}
\end{align}
where we used the definition of $\Sigma_{M}$ in Eq. \ref{eq:self_energy_as_minus_susceptibility}.
Collecting the results of Eqs. \ref{eq:AG0a+}, \ref{eq:AG0VG0a},
and \ref{eq:AGVGVGa} into Eq. \ref{eq:AGa+} (with the corresponding
superscripts “eff”), we can solve for $\mathcal{A}\mathcal{G}^{\text{eff}}a^{\dagger}$
to obtain (see Eq. \ref{eq:D_R_as_Geff}),

\begin{align}
D^{R}(\omega)= & \mathcal{A}\mathcal{G}^{\text{eff}}a^{\dagger}\nonumber \\
= & \mathcal{A}\mathcal{G}^{\text{eff}}a^{\dagger}\nonumber \\
= & \frac{1}{(\omega-\omega_{c}+i\frac{\kappa}{2})-\Sigma_{M}},\label{eq:aH_omega_harmonic_kubo}
\end{align}
which plugged into Eq. \ref{eq:a_omega_perturbative} yields Eq. \ref{eq:a_H(omega)_harmonic}.

To conclude, let us compare the calculation in \ref{subsub EoM} and
the current one, which give the same answers for $\langle a(\omega)\rangle$.
The latter assumes: (a) weak drive and (b) a stationary initial condition
$\rho(t_{\text{in}})$; while the former does not demand these restrictions.
However, both assumptions are unnecessary for harmonic systems \citep{mukamel2011quantum},
since their response to a driving field is (a) linear no matter how
strong the drive is, (b) independent of initial condition.

\bibliographystyle{ieeetr}
\bibliography{linear_response}

\end{document}